\documentclass[sn-aps,iicol]{sn-jnl}
\usepackage{graphicx} 
\usepackage{tikz}
\usetikzlibrary{arrows.meta} 
\usetikzlibrary{calc}
\usepackage{amsthm}
\usepackage{amsfonts}   
\usepackage{amsmath}   
\usepackage{enumitem}
\usepackage{subcaption}
\usetikzlibrary{matrix,positioning, backgrounds}
\usepackage{booktabs}
\usepackage{tabularx}
\usepackage{algorithm}
\usepackage{algorithmicx}
\usepackage[noend]{algpseudocode}
\usepackage[table]{xcolor}
\usepackage{csvsimple}
\usepackage{amssymb}
\usepackage{lineno}

\usepackage{booktabs}
\usepackage{multirow}
\usepackage{siunitx}
\usepackage{threeparttable}
\usepackage{adjustbox}
\usepackage{balance}

\usepackage{geometry}
\usepackage{pdflscape}
\usepackage{booktabs}
\usepackage{longtable}
\usepackage{array}
\usepackage[strict]{changepage}

\algrenewcommand\algorithmicrequire{\textbf{Input:}}
\algrenewcommand\algorithmicensure{\textbf{Output:}}

\title[Diagonal Packing for Efficient Homomorphic Sparse Matrix–Vector Multiplication]{Diagonal Packing for Efficient Homomorphic Sparse Matrix–Vector Multiplication}

\author[1]{\fnm{Kemal} \sur{Mutluergil}}\email{kemal.mutluergil@sabanciuniv.edu}
\author[1,2]{\fnm{Deniz} \sur{Elbek}}\email{deniz.elbek@sabanciuniv.edu} \author*[1,2,]{\fnm{Kamer} \sur{Kaya}}\email{kamer.kaya@sabanciuniv.edu}
\author[1]{\fnm{Erkay} \sur{Savaş}}\email{erkay.savas@sabanciuniv.edu}
\affil[1]{\orgdiv{Faculty of Engineering and Natural Sciences}, \orgname{Sabancı University}, \orgaddress{\city{Istanbul},  \country{Turkey}}}
\affil[2]{\orgdiv{VERIM, Center of Excellence in Data Analytics}, \orgname{Sabancı University}, \orgaddress{\city{Istanbul}, \country{Turkey}}}

\abstract{Homomorphic encryption (HE) enables computation over encrypted data but incurs a substantial overhead. For sparse matrix-vector multiplication, the widely used Halevi-Shoup scheme works over the non-empty diagonals, which may be many due to the irregular nonzero pattern of the matrix. 
Existing HE matrix-vector methods either use dense diagonal packing, which wastes rotations on empty diagonals, or sparse-coordinate compression, which can expose structural metadata. In this work, we instead keep the diagonal-method representation but reorder rows and columns to reduce the number of occupied cyclic diagonals. We formalize this problem as the {\em 2D-diagonal packing problem} and provide an integer programming formulation that yields optimal solutions for small instances. For large matrices, we propose practical {\em ordering} and iterative-improvement-based optimization heuristics.
We also introduce a dense row/column elimination strategy.
Experiments on 175 real-life matrices show that our ordering–optimization variants can reduce the diagonal count by $5.5\times$ on average ($45.6\times$ for one instance). In addition, the dense row/column elimination approach can be useful for cases where the proposed permutation techniques are not sufficient; for instance, in one case, the additional elimination helped to reduce the encrypted multiplication cost by $23.7\times$ whereas without elimination, the improvement was only $1.9\times$.}

\date{October 2025}

\newcommand{\Ab}{\mathbf{A}}
\newcommand{\Cb}{\mathbf{C}}
\newcommand{\Pb}{\mathbf{P}}
\newcommand{\Qb}{\mathbf{Q}}

\newcommand{\db}{\mathbf{d}}
\newcommand{\xb}{\mathbf{x}}
\newcommand{\yb}{\mathbf{y}}

\newcommand{\nnz}{\mathrm{nnz}}

\theoremstyle{definition}
\newtheorem{definition}{Definition}[section]

\begin{document}

\maketitle
\section{Introduction}

Homomorphic encryption (HE) allows computations to be performed on {\em ciphertexts} whose outputs can be decrypted to the same value as if they were performed on {\em plaintexts}. This enables outsourcing the computation for many applications, e.g., neural‑network inference and statistical analysis on sensitive data~\citep{Meftah2021, JoonWoo2022}. However, HE also has a significant computational overhead. 

Let $\Ab$ be an $n \times n$ matrix and $a_{i,j}$ be its entry at the $i$th row and $j$th column for $0 \leq i,j < n$. The matrix–vector multiplication is a fundamental block;
for instance, in a neural network, a single layer computes $\yb = \mathbf{A}\xb$ for column vectors $\xb$ and $\yb$. 
Let $\Ab$ be partitioned into $n$ non-overlapping, cyclic diagonals, where the diagonal $k$ for $0 \leq k < n$ contains the entries $a_{i, (k + i) \bmod n}$ for $0 \leq i < n$. If (at least) one entry in a diagonal is nonzero, it is called {\em non-empty}. For the encrypted matrix-vector multiplication, Halevi-Shoup packs entries along wrapping diagonals and rotates $\xb$ accordingly before each vector multiplication~\cite{halevishoup}. Hence, when $\Ab$ is dense, $n$ encrypted vector-vector multiplications are performed. 
When $\Ab$ is sparse, these operations can be omitted for empty diagonals. 

Existing HE matrix-vector methods either use dense diagonal packing, which wastes rotations on empty diagonals, or sparse-coordinate compression, which can expose structural metadata. In this work, {\em{we instead keep the diagonal-method representation but reorder rows and columns to reduce the number of occupied cyclic diagonals}}. 

Classical sparse matrix reordering optimizes memory locality; HE SpMV needs a different objective. In Halevi–Shoup-style SpMV, the expensive unit is not a cache miss or a nonzero, but an occupied cyclic diagonal, because each occupied diagonal induces rotations and encrypted vector operations. Therefore, sparse matrices with the same $n$ and number of nonzeros can have  different HE costs depending purely on their diagonal support. This paper formulates this new layout problem and studies practical algorithms for it. 

More formally, we focus on the following optimization problem: {\em Given a sparse matrix $\Ab$, find two permutation matrices ${\mathbf P}$ and ${\mathbf Q}$ such that ${\mathbf A'} = {\mathbf{PAQ^{\top}}}$ has as few non‑empty cyclic diagonals as possible}. The permutations ${\mathbf P}$ and ${\mathbf Q}$ reorder the rows and columns of ${\mathbf A}$, respectively. If $\yb = \mathbf{A}\xb$ then we have $\yb = \mathbf{P}^{\top}\mathbf{A'}\left({\mathbf Q}\xb\right)$. Hence, one can use ${\mathbf A'}$ instead of ${\mathbf A}$ to reduce the complexity of HE operations. Overall, the proposed approach should be viewed as a compiler-level optimization for homomorphic linear algebra: it changes neither the plaintext linear operator nor the HE scheme, but it reduces the number of expensive encrypted rotations and vector multiplications required by the diagonal method.

The main contributions of the manuscript are summarized as follows:
\begin{itemize}[leftmargin=*]
\item We identify occupied cyclic diagonals as the relevant cost metric
for Halevi-Shoup-style encrypted SpMV and formulate row-column reordering for
this objective as the two-dimensional diagonal packing problem (2DPP).

\item Second, we connect 2DPP to classical matrix layout notions, provide a lower bound
based on maximum row/column degree, and give a binary ILP formulation that can
certify optimal solutions on small instances.

\item Third, we propose a scalable heuristic framework that combines graph-ordering
initializers with 2OPT/3OPT local improvement moves evaluated directly on the
proposed objective.

\item Fourth, we show on 175 SuiteSparse matrices that the best combined variant obtains
the lowest diagonal count on 160 matrices and reduces the diagonal count by $5.5\times$  on
average over the natural ordering. We further show that dense row/column elimination
addresses a separate obstruction and can produce large additional runtime reductions.
\end{itemize}

The rest of the manuscript is organized as follows: Section~\ref{sec:nota} introduces the notation and background, and Section~\ref{sec:2dpp} formalizes the 2D Diagonal Packing Problem and discusses how the resulting permutations can be deployed and shared in practical HE scenarios. Section~\ref{sec:algorithms} describes the ordering heuristics used together with the iterative-improvement scheme. The dense row/column elimination and the associated HE cost model are provided in Section~\ref{sec:dense}. Section~\ref{sec:exp} reports the experimental results, and Section~\ref{sec:review} reviews related work. Finally, Section~\ref{sec:conc} concludes the paper and outlines future research directions.

\section{Notation and Background}\label{sec:nota}

A \emph{permutation matrix} $\Pb\in\{0,1\}^{n\times n}$ has exactly one~1 in each row/column and satisfies $\Pb^{-1}=\Pb^{\top}$. 
Each one-to-one mapping $\pi:\{1,\dots,n\}\!\to\!\{1,\dots,n\}$ corresponds
to a permutation matrix $\mathbf{P}_\pi$ with $n$ entries $p_{i, \pi(i)} = 1$, and 0 elsewhere. For a symmetric, $n \times n$ matrix $\mathbf{A}$, the permutation is applied to the rows and columns as $\mathbf{A}_\pi \;=\; \mathbf{P}_\pi\,\mathbf{A}\,\mathbf{P}_\pi^{\top}$.
The \emph{width} of the mapping $\pi$ is 
\[
w(\Ab_\mathbf{\pi}) \;=\; \max_{\,a'_{i,j} {\mbox{\footnotesize{ is a nonzero in }}} \Ab_\pi}\, |i-j| ,
\]
whereas the \emph{bandwidth} of $\Ab$, $\operatorname{bw}(\mathbf{\Ab}) \;=\; \min_{\pi} \; w(\Ab_\pi)$,
is the smallest width across all $\pi$.

Let $s(\Ab_\pi)$ be the number of distinct index differences over all the nonzeros of $\mathbf{A}_\pi$, i.e., 
$\bigl|\{\,|i-j|:\; a'_{ij} {\mbox{ is nonzero in }} \Ab_\pi\}\bigl|$. Then the \emph{bandsize}~\citep{ERDOS1988117,HeinrichHell1987Bandsize} of $\Ab$ is defined as
$
\operatorname{bs}(\mathbf{A}) \;=\; \min_{\pi}\; s(\Ab_\pi).
$
In particular, $\operatorname{bs}(\mathbf{A}) \le \operatorname{bw}(\mathbf{A}) + 1$ since, for any mapping, the realized differences form a set of integers contained in $\{0, 1, \ldots, w(\Ab_\mathbf{\pi})\}$. For a symmetric matrix $\Ab$, deciding if $\operatorname{bs}(\Ab) < k$ is an NP-complete problem for every fixed $k \geq 2$~\citep{Sudborough1986Bandsize}.

A binary matrix $\mathbf{C}\in\{0,1\}^{n\times n}$ is \emph{circulant} if each row is obtained via a cyclic right shift of the preceding row. If $\mathbf{C}$ is circulant, the whole matrix can be specified by a single set ${\cal C}_\Cb = \{c_0, c_1, c_2, \ldots, c_\ell\}$, which is the set of column indices of the nonzeros in its first row. For a symmetric, square matrix $\Ab$, Erd\"{o}s et al. introduced the {\em circular bandsize}, $\operatorname{cbs}(\Ab)$\footnote{Erd\"{o}s et al. leveraged {\em circular bandwidth}; {\em circular bandsize} is introduced only for symmetry.}, as the smallest possible $\ell$ such that the nonzero pattern of $\mathbf{P}_\pi\,\mathbf{A}\,\mathbf{P}_\pi^{\top}$ is contained in that of a symmetric circulant matrix $\Cb$ with $\ell =|{\cal C}_\Cb|$~\cite{ERDOS1988117}.

Let $\Ab \in {\mathbb R}^{n \times n}$ be an $n \times n$ matrix. In Halevi-Shoup matrix vector multiplication, $\Ab$ is partitioned into $n$ non-overlapping, cyclic diagonals, each containing $n$ entries. The vector for diagonal $k$, $0 \leq k < n$, contains the entries $\db_k = \{a_{i, (k + i) \bmod n}: 0 \leq i < n\}$ 
in increasing $i$ order. Hence, each matrix entry $a_{i,j}$ resides at the $i$th location of a nonzero vector $\db_k$ where $k = (j-i) \bmod n$. 

Let $\xb\in\mathbb{R}^{n}$ be the input vector and $\yb=\Ab\xb\in\mathbb{R}^{n}$ be the vector to be computed. 
Using $\mathrm{rot}^k(\xb)$ for the cyclic left-rotation of $\xb$ by $k$ (mod $n$) and $\odot$ for component-wise, i.e., Hadamard, product, the $\yb$ vector can be computed as
\begin{equation}
\yb
\;=\;
\sum_{k=0}^{n-1} \db_k \odot \mathrm{rot}^k(\xb)\label{matvec}
\end{equation}
\cite{halevishoup} leverages~Eq.~\ref{matvec} albeit with encrypted inputs. Let $Enc$ be a homomorphic encryption function and $Enc(\db_k)$, $Enc(\xb)$, $Enc(\yb)$ be the {\em ciphertext}s, respectively, for $\db_k$, $\xb$ and $\yb$. The operations $\mathrm{rot}$ and $\odot$, as well as the additions within $\sum$, are efficiently implemented by state-of-the-art HE libraries to perform 
\begin{equation}   \label{eq:diag-decomp}
 Enc(\yb) 
 \;=\; 
 \sum_{k=0}^{n-1} Enc(\db_k) \odot \mathrm{rot}^k(Enc(\xb)).
\end{equation}
Although both the matrix~(diagonals) and the vector $\xb$ are encrypted in~\eqref{eq:diag-decomp}, in practice, only one of them can be encrypted based on the use case. 

\section{The 2DPP Problem}\label{sec:2dpp}

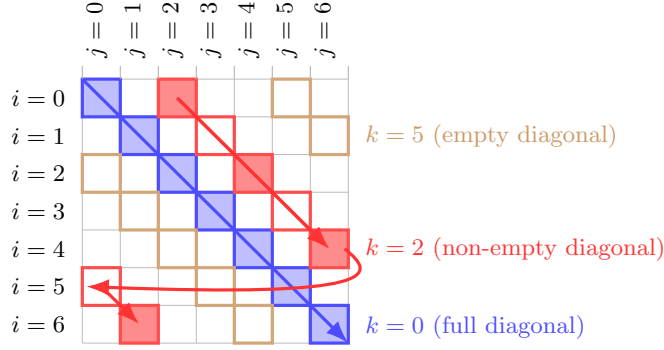
\begin{figure}[htbp]
\centering
\begin{tikzpicture}[scale=0.44, every node/.style={font=\small}]
  \def\n{7}
  \def\labelgap{1.3}

  \begin{scope}[yscale=-1, every node/.append style={transform shape=false}]
    \fill[white] (0,0) rectangle (\n,\n);

    \foreach \i in {0,...,\numexpr\n} {
      \draw[gray!50] (0,\i) -- (\n,\i);
      \draw[gray!50] (\i,0) -- (\i,\n);
    }

    \foreach \i in {0,...,\numexpr\n-1} {
      \fill[blue!25] (\i,\i) rectangle ++(1,1);
       \draw[blue!70, line width=1.2pt] (\i,\i) rectangle ++(1,1);

      \node at (\i+0.5,\i+0.5) {};
    }
    \draw[blue!70, line width=1.2pt, -{Latex[length=3mm]}] (0,0) -- (\n,\n);

    \foreach \p/\q in {2/0,4/2,6/4,1/6} {
      \fill[red!45] (\p,\q) rectangle ++(1,1);
      \draw[red!70, line width=1.2pt] (\p,\q) rectangle ++(1,1);
      \node at (\p+0.5,\q+0.5) {};
    }

    \foreach \p/\q in {5/0,6/1,0/2,1/3,2/4,3/5,4/6} {
      \draw[brown!70, line width=1.2pt] (\p,\q) rectangle ++(1,1);
      \node at (\p+0.5,\q+0.5) {};
    }

    \foreach \p/\q in {3/1,5/3,0/5} {
      \fill[white!25] (\p,\q) rectangle ++(1,1);
      \draw[red!70, line width=1.2pt] (\p,\q) rectangle ++(1,1);
      \node at (\p+0.5,\q+0.5) {};
    }
    
    \draw[red!80, line width=1.2pt, -{Latex[length=3mm]}]
      (2.5,0.5) -- (3.5,1.5) -- (4.5,2.5) -- (5.5,3.5) -- (6.5,4.5);
    \draw[red!80, line width=1.2pt, -{Latex[length=3mm]}] (2.5,0.5) -- (3.5,1.5) -- (4.5,2.5) -- (5.5,3.5) -- (6.5,4.5); \draw[red!80, line width=1.2pt, -{Latex[length=3mm]}] (6.9,4.5) .. controls (7.8,5.2) and (7.8,5.8) .. (0.1,5.5); \draw[red!80, line width=1.2pt, -{Latex[length=3mm]}] (0.5,5.5) -- (1.5,6.5);

    \foreach \i in {0,...,\numexpr\n-1} {
      \node[left] at (-0.2,\i+0.5) {$i=\i$};
    }

    \foreach \j in {0,...,\numexpr\n} {
      \draw[gray!60] (\j,0) -- (\j,-0.35);
    }
    \foreach \j in {0,...,\numexpr\n-1} {
      \node[rotate=90, above] at (\j+0.9,-\labelgap) {$j=\j$};
    }

    \node[blue!70, anchor=west] at (\n+0.2,6.5\n) {$k=0$ (full diag.)};
    \node[red!80,  anchor=west] at (\n+0.2,4.5) {$k=2$ (non-empty diag.)};
    \node[brown!80,  anchor=west] at (\n+0.2,1.5) {$k=5$ (empty diag.)};

  \end{scope}
\end{tikzpicture}
\vspace{2ex}
\caption{A $7\times 7$ matrix with 11 nonzeros (filled squares) on 2 diagonals $\db_0$ and $\db_2$.}
\label{fig:diags}
\end{figure}

For a {\em sparse matrix} $\Ab$, one can omit the multiplications, rotations, and additions in Eq.~\ref{eq:diag-decomp} of the Halevi-Shoup scheme if the corresponding diagonals are {\em empty}. Fig.~\ref{fig:diags} shows a $7 \times 7$ matrix with $\tau = 11$ nonzeros~(filled squares), and with $2$ non-empty diagonals. In the figure, 
the main diagonal $\db_0$ is {\em full}, i.e., its entries are nonzeros, $\db_2$ is {\em non-empty} with 4 nonzeros, and $\db_5$ is {\em empty} with all 0s similar to $\db_1$, $\db_3$, $\db_4$ and $\db_6$. 

 When $\Ab$'s nonzero pattern is circulant,  
 each integer $c \in {\cal C}_{\Ab}$ maps to a non-empty diagonal $\db_{c}$. 
When $|{\cal C}_\Ab|$ is small compared to $n$, Eq.~\ref{eq:diag-decomp}, designed for dense matrices, is extremely inefficient since an empty $\db_k$ will not have an impact on the final result. In fact, this is not only true for circulant matrices but for all matrices $\Ab$ with a small $\operatorname{cbs}(\Ab) \ll n$. To avoid such inefficiencies, \eqref{eq:diag-decomp} must be rewritten as 
\begin{equation}\label{eq:diag-decomp-nonzero}
Enc(\yb)
\;=\;
\sum_{k \in \mathcal{D}_\Ab} Enc(\db_k) \odot \mathrm{rot}^k(Enc(\xb)).
\end{equation}

The {\em bandsize} and {\em circular bandsize} are  
defined for (undirected) graphs; hence,  
the literature focuses on symmetric matrices.  
However, in general for {SpMV}, we do not have that restriction. Furthermore, the rows and columns of the matrix can be permuted via different permutations. In this work, we focus on the following problem:

\begin{definition}[2D Diagonal Packing Problem~({\bf{2DPP}})]
Given a sparse, $n \times n$ matrix $\Ab$, find two permutations $\pi_R$ and $\pi_C$ that minimize $s(\Ab_{\pi_R, \pi_C})$ which equals to
\begin{small}
\begin{align*}
|\{k \in {\mathbb Z}_n : &\exists a_{i,j}  \mbox{ s.t. } k = (\pi_C(j) - \pi_R(i)) \bmod n\}|\
\end{align*}
\end{small}
\noindent which is the number of non-empty cyclic diagonals for the matrix $\Ab' = \Pb_{\pi_R}\Ab\Qb^{\top}_{\pi_C}$.
\end{definition}
 
Let $\operatorname{cbs_{2D}}(\Ab)$ be {\em{2D circular bandsize}}, i.e., the minimum over all possible $\pi_R$, $\pi_C$:
\[
\operatorname{cbs_{2D}}(\Ab)
\;=\;
\min_{\pi_R,\pi_C}\; s(\Ab_{\pi_R, \pi_C}).
\] 
Let $\delta_R(i) = |\{j: a_{i,j}\neq 0\}|$ and $\delta_C(j) = |\{i: a_{i,j}\neq 0\}|$ be the {\em degrees}, i.e., the number of nonzeros, of row $i$ and column $j$, respectively. Then, since each nonzero in the same row or column lies in a different diagonal for any permutation, 
\begin{equation}
\label{eq:bound}
{\operatorname{cbs_{2D}} \;\ge\; \max\left\{\max_{0 \leq i < n}{\delta_R(i)},\max_{0 \leq j < n}{\delta_C(j)}\right\}}.
\end{equation}

\subsection{ILP Formulation of 2DPP}
The binary Integer Linear Programming formulation of 2DPP has the following \emph{variables}: 
\begin{itemize}[noitemsep]
\item $p_{i,i'}\in\{0,1\}$: row $i$ gets position $i'$ \\for $0 \leq i, i' < n$,
\item $q_{j,j'}\in\{0,1\}$: column $j$ gets position $j'$ \\for $0 \leq j, j' < n$,
\item $z_k\in\{0,1\}$: cyclic diagonal $k$ is used \\for $0 \leq k < n$,
\end{itemize}
\noindent where the \emph{objective} function is 
\[
\min \;\sum_{0 \leq k < n} z_k,
\]
\noindent under the following \emph{permutation constraints}
\begin{align}
&\sum_{0 \leq i' < n} p_{i,i'} = 1 && \mbox { for }0 \leq i < n, &\\
&\sum_{0 \leq i < n} p_{i,i'} = 1 && \mbox { for }0 \leq i' < n, &\\
&\sum_{0 \leq j' < n} q_{j,j'} = 1 && \mbox { for }0 \leq j < n, &\\
&\sum_{0 \leq j < n} q_{j,j'} = 1 && \mbox { for }0 \leq j' < n. &
\end{align}

\noindent and the \emph{diagonal activation constraint}
\[
z_{((j'-i') \bmod n)} \;\;\ge\;\; p_{i,i'} + q_{j,j'} - 1
\]
for all nonzero $a_{i,j} {\mbox{ and }} 0 \leq i', j'< n$. In practice, the symmetry can be broken by fixing one permutation
entry, e.g., $p_{0,0} = 1$.

\subsection{Using the 2DPP permutations}
Let the permutations
$\Pb_{\pi_R}$ and $\Qb_{\pi_C}$ be used to minimize the number of non-empty cyclic diagonals in
$\Ab'=\Pb_{\pi_R}\,\Ab\,\Qb^{\top}_{\pi_C}$. With $\Ab'$, there are three steps to compute $\yb$: 

\noindent
\begin{tabular}{l|p{2.5in}}   
1) $ \xb' =\Qb_{\pi_C}\xb$ & {\em preprocess}: to permute the en-\\ 
&tries of the input vector,\\
2) $\yb'=\Ab'\xb$ & {\em compute}: the actual expensive\\ 
&SpMV product,\\
3) $\yb =\Pb_{\pi_R}^{\top}\yb'$ & {\em postprocess}: to put the output\\ &entries to correct places.
\end{tabular}

We outline three common deployment patterns; in each case, $\Ab$ is assumed to correspond to a {\em model}, and its owner generates $(\Pb_{\pi_R},\Qb_{\pi_C})$, and $\Ab'$ in plaintext form.\\

\noindent{\bf{\underline{Scenario 1} - Encrypted matrix (model to client)}:} The model owner encrypts $\Ab'$ and sends it to the client. 
She also shares $\Qb_{\pi_C}$ so the client can form $\xb'=\Qb_{\pi_C}\xb$. The client evaluates $Enc(\yb')=Enc(\Ab')\xb'$ using HE-supported plaintext-ciphertext primitives as available, sends $Enc(\yb')$ back to the model owner, which is first decrypted and then repermuted via $\yb=\Pb_{\pi_R}^{\top}\yb'$ and sent back to the client.\\

\noindent{\bf{\underline{Scenario 2} - Encrypted vector (client to model):}} 
The model owner sends $\Qb_{\pi_C}$ to the client, who computes $\xb'=\Qb_{\pi_C}\xb$,
for a private query $\xb$, encrypts it as $Enc(\xb')$, and sends it to the model owner. The owner 
performs the plaintext-ciphertext product $Enc({\mathbf y'}) = \Ab' Enc(\xb')$, and returns $Enc(\yb')$ and $\Pb_{\pi_R}$ to the client.\\

\noindent{\bf{\underline{Scenario 3} - Both sides encrypted (outsourced server):}} A third–party server evaluates the HE computation. The model owner provides
$(\Pb_{\pi_R},\Qb_{\pi_C})$ to coordinate preprocessing: the client forms and encrypts
$\xb'=\Qb_{\pi_C}\xb$, the owner encrypts $\Ab'$, and the server computes
$Enc(\yb')=Enc(\Ab')Enc(\xb')$ via ciphertext-ciphertext primitives. Either can apply $\Pb_{\pi_R}^{\top}$.\\

In {\em Scenario~2}, the client learns only the permutations $\Pb_{\pi_R}$ and $\Qb_{\pi_C}$, not which diagonals are occupied by $\Ab'$. In {\em Scenarios~1} and~{\em{3}}, the server/client may infer the \emph{set} of diagonal indices
of $\Ab'$ but not the positions of nonzeros within the diagonals. 
This residual leakage can be further reduced by using extra, all-zero diagonals or by randomizing diagonals across runs. 

In practice, 2DPP optimizes a simplified surrogate of the exact HE cost. Let $S$
denote the number of slots available in a ciphertext vector~($S = 4096$ in our Cheon–Kim–Kim–Song~(CKKS) implementation). When $n \leq S$, each
cyclic diagonal can be stored in a single ciphertext, and minimizing the number of
non-empty diagonals is equivalent to minimizing the number of ciphertext vectors that
must be processed. When $n > S$, however, a single diagonal no longer fits into one
ciphertext and must be split across
$\left\lceil n / S \right\rceil$ ciphertexts. Hence, for large $n$, minimizing the number of diagonals is not identical to minimizing the exact
number of ciphertext vectors. 
However, our experimental results indicate that this approximation is an effective surrogate and is highly useful in practice.

\section{Iterative-Improvement Heuristics for 2DPP}~\label{sec:algorithms}

To reduce the cost of an encrypted SpMV, we propose an iterative-improvement-based approach for 2DPP, which, given $\Ab$, (1) finds a good initial ordering pair $\pi_R$ and $\pi_C$ to pack its nonzeros into a small number of non-empty diagonals, and (2) iteratively refines these permutations for a much smaller $s(\Ab_{\pi_R, \pi_C})$. The speedup achieved by this approach is proportional to the reduction in the number of non-empty diagonals. 
For instance, the matrix {\tt{Luxemburg OSM}}, with $n = 114,599$  
and $119,666$ nonzeros\footnote{https://sparse.tamu.edu/}, has 21,866 diagonals in its {\em natural} order. 
The proposed approach reduces it to 483~(45.3$\times$ improvement), and 
the encrypted SpMV takes 180 seconds, whereas originally, it takes two hours~(i.e., $40\times$ speedup).

\subsection{Initializing $\pi_R$ and $\pi_C$}

To the best of our knowledge, there is no algorithmic work focusing on 2DPP. However, as Section~\ref{sec:2dpp} explains, the problem has strong connections to the literature. To handle the first step that generates the initial permutations $\pi_R$ and $\pi_C$, we relate 2DPP to three concepts; {\em bandwidth}, {\em anti-bandwidth}, and {\em circulant graphs}. 

The ordering heuristics adopted from the literature assume a symmetric
matrix. For a generic $\Ab \in \mathbb{R}^{n \times n}$, let $\mathbf{B}$ be the binary matrix with the same nonzero pattern, i.e., $b_{i,j} = 1$ if and only if $a_{i,j} \neq 0$. We leverage two symmetrization constructions: 
We use the sparsity patterns of {(1)} $\mathbf{\hat{B}} = \mathbf{B} + \mathbf{B}^\top$ and {(2)} 
$\mathbf{\hat{B}} = \begin{bmatrix}
\mathbf{0} & \mathbf{B} \\
\mathbf{B}^\top & \mathbf{0}
\end{bmatrix}.$ 

For symmetrization with $\mathbf{\hat{B}} = \mathbf{B} + \mathbf{B}^\top$, the ordering $\pi^\star$ on $\mathbf{\hat{B}}$ is used to initialize both $\pi_R = \pi_C = \pi^\star$. On the other hand, for the latter, the ordering yields a $\pi^\star$ for $2n$ nodes. From $\pi^\star$, we extract 
$\pi_R$ and $\pi_C$ by preserving the relative order of the $n$, respectively, row- and column-nodes in $\pi^\star$. This construction directly produces \emph{distinct} row and column permutations, which better reflect the freedom allowed by the 2DPP problem.

\subsubsection{The Bandwidth and Reverse Cuthill--McKee Ordering}

The circular bandsize is related to the regular bandsize; since each $|i - j|$ can map to two different residues in modulo $n$, i.e., $\operatorname{cbs_{2D}}(\Ab) \le 2 \times \operatorname{bs}(\Ab)$.
Moreover, we know that $\operatorname{bs}(\Ab) \le \operatorname{bw}(\Ab) + 1$~\citep{ERDOS1988117};  hence, $\operatorname{cbs_{2D}}(\Ab) \le 2 \times (\operatorname{bw}(\Ab) + 1)$. In light of these, we 
leverage a bandwidth reduction heuristic, {\em Reverse Cuthill-McKee}~(RCM)~\citep{10.1145/800195.805928}, to obtain the initial orderings $\pi_R$ and $\pi_C$. 

RCM reorders the rows/columns of a sparse {\em symmetric} matrix $\Ab$ to concentrate the nonzeros around the main diagonal. It operates on the graph $G_\Ab = (V, E)$ with $n$ vertices, where each vertex $i$ corresponds to a $i$th row/column, and the edges connect vertex pairs $(i,j) \in E$ for which $a_{ij} \neq 0$. The algorithm first chooses a \emph{pseudo-peripheral node}, i.e., one that is far from the center of the graph, using Alg.~\ref{alg:ppn-high}. It then performs a BFS while visiting the vertices in the order of increasing degrees. The visit order in the final BFS is reversed and used as the output permutation. 

For disconnected graphs, we apply the procedure component-wise: after the current component is exhausted, we choose a new pseudo-peripheral root among the unlabeled vertices and continue; component order is determined by decreasing component size.

\begin{algorithm}[htbp]
\small
\caption{$u = $ {\sc PseudoPeripheral}($G_\mathbf{\hat{B}}$)}
\label{alg:ppn-high}
\begin{algorithmic}[1]
\Require Graph $G_\mathbf{\hat{B}}$ of the symmetrized matrix $\mathbf{\hat{B}}$
\Ensure Pseudo-peripheral node $u \in G_\mathbf{\hat{B}}$
\State $u \gets$ a random node in $V$ 
\State $\textit{bestDepth} \gets 0$
\State $\textit{streak} \gets 0$
\While{$\textit{streak} < T$}\label{ln:bfsloop} \Comment{$T$ is small (e.g., 5--10)}
  \State Run BFS from $u$ 
  \State Obtain level sets $L_0 = \{u\},L_1,\dots,L_{d-1}$ 
  \If{$d > \textit{bestDepth}$}
    \State $\textit{bestDepth} \gets d$;
    \State $\textit{streak} \gets 0$
  \Else
    \State $\textit{streak} \gets \textit{streak} + 1$
  \EndIf
  \State ${u} \gets$ random vertex in the farthest level $L_{d-1}$ 
\EndWhile
\State \Return ${u}$
\end{algorithmic}
\end{algorithm}

\subsubsection{The Anti-bandwidth and Miller-Pritikin Ordering}

For a {\em symmetric} sparse matrix, the
\emph{anti-width} of a mapping $\pi$ is 
\[
aw(\Ab_\mathbf{\pi}) \;=\; \min_{\,a'_{ij} {\mbox{\footnotesize{ is a nonzero in }}} \Ab_\pi}\, |i-j| ,
\]
and the \emph{anti-bandwidth} of $\Ab$, $
\operatorname{abw}(\mathbf{\Ab}) \;=\; \max_{\pi} \; aw(\Ab_\pi)
$ is the largest anti-width over all possible mappings.
$\operatorname{abw}$ maximizes the minimum nonzero distance to the main diagonal across all nonzeros. This problem is the ``dual'' of
bandwidth minimization and is NP-complete~\citep{leung1984some}. For mesh graphs, bounds and constructions
exist~\citep{miller1989separation,raspaud2009antibandwidth}. 

Let $u$ be a corner vertex in the rectangular grid $G = (V,E)$ of size $m\times k$. The {\sc Miller-Pritikin}~(MP) algorithm initiates BFS from $u$ to construct levels $L_i=\{v\in V:d(u,v)=i\}$ for $i\in\{0,1,\dots,m+k-2\}$. On the grid, edges only join consecutive levels, i.e., every edge in $E$ has one endpoint in $L_i$ and the other in $L_{i\pm 1}$. Consequently, if $i\equiv j\pmod 2$, there is no edge between any vertex in $L_i$ and any vertex in $L_j$. Motivated by this fact,~\cite{miller1989separation} orders the vertices level by level in two parity blocks: (1) the even levels $L_{even} = (L_0, L_2,\dots)$ and (2) odd levels $L_{odd} =(L_1, L_3,\dots)$, so the final permutation $\pi^{\star}_{mp}$ is obtained by concatenating $L_{even}$ and $L_{odd}$. Within each level, the vertices are ordered arbitrarily, as the grid structure ensures that there are no intra-level edges.  
Although this algorithm is optimal on grid graphs, it still performs well on general graphs, as detailed in Section~\ref{sec:exp}.

\begin{algorithm}[htbp]
\small
\caption{$\pi^*_\text{lb} = $ {\sc LBS}($G_\mathbf{\hat{B}} = (V,E)$, $u$)}
\label{alg:lb-antibw}
\begin{algorithmic}[1]
\Require Graph $G_\mathbf{\hat{B}} = (V,E)$ of $\mathbf{\hat{B}}$, \newline
\hspace*{4.9ex}A BFS root vertex $u \in G_\mathbf{\hat{B}}$ with level \newline \hspace*{4.5ex} sets $L_0 = \{u\},L_1,\dots,L_{d-1}$
\Ensure Mapping $\pi^*_{lb}:V\!\to\!\{0,\dots,n-1\}$
\State $\pi^{*}_{lb}(u) \leftarrow 0$;
\State $i \gets 1$;
\State $\textit{sweep} \gets 0$;
\State $\textit{flag}(v)\gets 0$ for all $v\in V$
\While{$i < n$}
  \State $\textit{sweep} \gets \textit{sweep}+1$
  \For{$r = 1$ \textbf{to} $d-1$}
    \ForAll{unlabeled $v \in L_r$}
      \If{$\textit{flag}(v) = \textit{sweep}$}
        \State \textbf{continue} \Comment{skip flagged vertices}
      \EndIf
      \State $\pi^*_\text{lb}(v) \gets i$
      \State $i \gets i+1$; 
      \ForAll{unlabeled neighbors $w$ of $v$}
        \State $\textit{flag}(w) \gets \textit{sweep}$
      \EndFor
    \EndFor
  \EndFor
\EndWhile
\State \Return $\pi^*_\text{lb}$
\end{algorithmic}
\end{algorithm}

Following the ideas of~\cite{miller1989separation}, Scott and Hu proposed Algorithm~\ref{alg:lb-antibw} that extends the {\sc Miller-Pritikin} algorithm from grid graphs to general graphs~\cite{https://doi.org/10.1002/nla.1859}. The key principle is to postpone assigning labels to neighbors of already labeled vertices as late as possible.
Given a root $u$ and its BFS level sets $L_0 = \{u\},L_1,\dots,L_{d-1}$, Algorithm~\ref{alg:lb-antibw}, {\sc{LBS}}, generates a permutation in multiple \emph{sweeps}. Within each sweep, it scans levels $1$ to $d-1$, and when an unlabeled vertex $v \in L_r$ is labeled, all of its currently unlabeled neighbors are \emph{flagged} so they will be skipped for the remainder of that sweep. This spreads consecutively assigned labels far apart, thereby increasing the minimum label distance across edges. The sweeps repeat until all vertices are labeled and hence a full permutation is obtained. When the underlying graph is a grid, the permutation $\pi^*_\text{lb}$ of Algorithm~\ref{alg:lb-antibw} is the same as $\pi^\star_{mp}$. 
Similar to RCM, we used the two symmetrization techniques mentioned in the previous subsection as well as Algorithm~\ref{alg:ppn-high} used to find the first vertex $u \in V$.

\begin{figure}
\centering
\includegraphics[width=\linewidth]{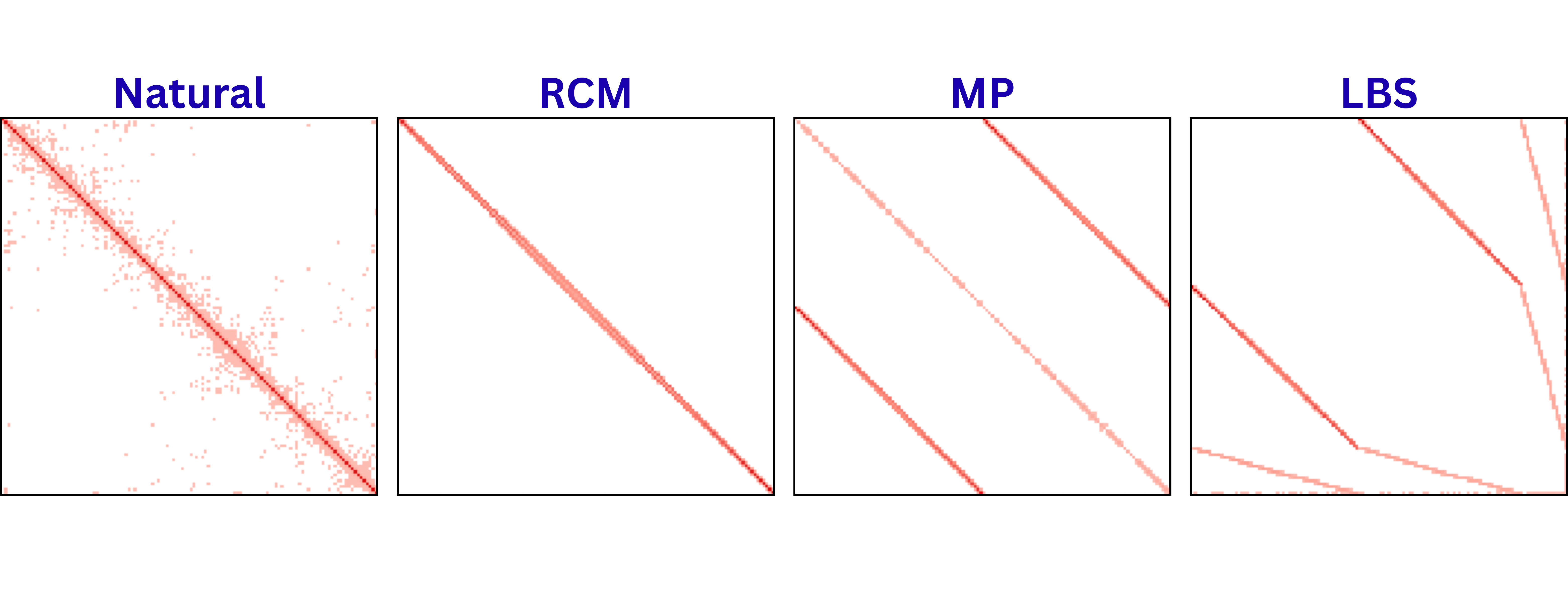}
\caption{Visualisation of the matrix {\tt{big\_dual}} in its natural form and after RCM, MP, and LBS.}
\label{fig:big_dual}
\end{figure}

To provide insight into how these combinatorial algorithms work, Figure~\ref{fig:big_dual} visualizes the permuted forms of the matrix {\tt big\_dual} having $n = 30,269$ rows/columns and $89,858$ nonzeros ordered by the combinatorial algorithms introduced in this section. 

\subsubsection{The Circulant Layout and Richter-Rocha Algorithm}
\label{sec:eigen}

For a given $\Ab$, 2DPP focuses on minimizing the number of distinct residues $(\pi_{C}(j)-\pi_{R}(i)) \bmod n$. When $\Ab$'s sparsity pattern is (almost) \emph{circulant}, Richter and Rocha demonstrated that its symmetric graph exhibits smooth, low-frequency eigenmodes~\cite{RichterRocha2017}. In their theoretical model, the eigenvectors corresponding to the 2nd and 3rd largest eigenvalues place vertices on a circle, with the polar angle correlating with the latent cyclic coordinate. This fits perfectly with the 2DPP problem, as we fit the randomly dispersed nonzeros of a matrix to a few circular diagonals. Hence, {\sc EigenOrder} can be a strong initializer for 2DPP at the cost of extracting a small set of top eigenvectors and performing a fast combinatorial evaluation. However, the noise in real-life matrices significantly perturbs the eigenspectrum. Since it is closely related to 2DPP, and the current treatment is too sensitive to noise, we keep the content in the Appendix.

\subsection{Iterative–Improvement Heuristics for 2DPP}
\label{sec:optimizer}

The second phase of the proposed approach takes the row/column permutations $\pi_R$ and $\pi_C$ produced by the heuristics above and \emph{refines} them by repeatedly applying perturbations. Each perturbation proposes a nearby configuration—ranging from small, local tweaks to larger, block-level rearrangements. While doing so, we keep track of the objective value $s(\Ab_{\pi_R,\pi_C})$. The {\em incumbent}, i.e., the solution at hand, is updated only when a proposal strictly reduces the number of occupied cyclic diagonals. This scheme therefore performs a biased exploration of the search space, favoring improving moves while still probing diverse neighborhoods through randomization.

A long line of work reduces the matrix \emph{bandwidth} and \emph{profile}/\emph{wavefront} via local, iterative improvement, e.g.,~\citep{https://doi.org/10.1002/nla.1859,GONZAGADEOLIVEIRA2020106434,10.1137/S1064827500379215,10.1145/592843.592844}. These methods all rely on small, local perturbations that monotonically improve a chosen objective (or accept ties according to secondary rules). As stated above, we focus on 
$s(\Ab_{\pi_R, \pi_C})$ defined in Def. 3.1.

Unlike bandwidth (a {\em maximum}) or antibandwidth (a {\em minimum})—both change by 1 when two {\em neighboring} rows/columns are swapped, $s(\Ab_{\pi_R, \pi_C})$ can change by several units after a single neighbor perturbation. Let rows \(r\) and \(r{+}1\) have degrees
\(\delta_R(r)\) and \(\delta_R({r+1})\). Suppose each nonzero in these two rows currently lies
on a diagonal that \emph{no other} row uses (worst case), and that after swapping
the two rows, the resulting diagonal indices are all \emph{new} with respect to
the rest of the matrix. Hence, one adjacent row/column swap can empty up or occupy up to
\(\Delta = \delta_R(r) + \delta_R(r+1)\)  
diagonals, so $s(\Ab_{\pi_R, \pi_C})$ may increase or decrease by \(\Delta\).
This makes the optimization process much more erratic and harder to manage compared to bandwidth/antibandwidth.

Given the permutations $(\pi_R,\pi_C)$ at hand, we maintain an array
$\texttt{d\_nnz}$ of size $n$ counting the number of nonzeros on each cyclic diagonal
$k\in\{0,\dots,n{-}1\}$ of $\Ab_{\pi_R,\pi_C}$. The \emph{primary} objective is minimizing 
the number of occupied diagonals
\begin{small}
\begin{align}
\#\text{diags} \;=\; \bigl|\{k:\texttt{d\_nnz}[k] > 0\}\bigr|,
\end{align}
\end{small}
which proxies $s(\Ab_{\pi_R,\pi_C})$. Our preliminary experiments reveal that only accepting operations that improve the {\em primary gain} on \#\text{diags} significantly hinders progress towards a local minimum. Hence, as \emph{secondary} guidance, we track
\begin{small}
\begin{align}
\texttt{min\_nnz} &\;=\; \min_{k:\,\texttt{d\_nnz}[k]>0} \texttt{d\_nnz}[k],
\\
\texttt{min\_nnz\_cnt} &\;=\; \bigl|\{k:\texttt{d\_nnz}[k]=\texttt{min\_nnz}\}\bigr|.
\end{align}
\end{small}
Reducing \texttt{min\_nnz} (and, if tied, increasing \texttt{min\_nnz\_cnt})
creates \emph{fragile} diagonals that are easier to eliminate in later steps.
In our implementation, a candidate move is accepted if it ({\bf i}) strictly decreases $\#\text{diags}$, or
({\bf ii}) keeps $\#\text{diags}$ the same but decreases \texttt{min\_nnz}, or
({\bf iii}) keeps both unchanged but increases \texttt{min\_nnz\_cnt}.
The possible moves considered in this work are described in Figure~\ref{fig:compact-moves} and summarized below:

\begin{itemize}[leftmargin=*]
    \item \textbf{2OPT:} Two rows (columns) are randomly selected, and their positions are exchanged. Permutations are adjusted slightly.

    \item \textbf{3OPT:} Three rows (columns) are randomly chosen, and their positions are shifted cyclically. Slightly more adjustment compared to 2OPT. 
\end{itemize}

\begin{figure}[htbp]
\centering
\newcommand{\DrawGrid}{%
  \foreach \i in {0,...,6} {\draw[gray!50] (0,\i) -- (6,\i);}
  \foreach \j in {0,...,6} {\draw[gray!50] (\j,0) -- (\j,6);}
}
\newcommand{\NZ}[1]{%
  \foreach \r/\c in {#1} {%
    \fill[black!75] (\c-1,6-\r) rectangle ++(1,1);
  }%
}
\newcommand{\HLrow}[2]{
  \fill[#2!25] (0,6-#1) rectangle (6,7-#1);
  \draw[#2!70, line width=0.6pt] (0,6-#1) rectangle (6,7-#1);
}

\def\A{1/5, 2/2, 2/6, 3/3, 4/1, 4/4, 6/2}

\begin{subfigure}{0.45\linewidth}
\centering
\begin{tikzpicture}[scale=0.25]
\begin{scope}
  \DrawGrid
  \HLrow{2}{red}\HLrow{4}{blue}
    \NZ{1/2, 2/5, 4/5, 4/3, 5/6}
\end{scope}
\draw[->,thick] (6.5,3) -- (7.5,3);
\begin{scope}[xshift=8cm]
  \DrawGrid
  \HLrow{2}{blue}\HLrow{4}{red}
  \NZ{1/2, 4/5, 2/5, 2/3, 5/6}
\end{scope}
\end{tikzpicture}
\caption{2OPT: exchange two rows (3$\rightarrow$2).}
\end{subfigure}\hfill
\begin{subfigure}{0.45\linewidth}
\centering
\begin{tikzpicture}[scale=0.25]
\begin{scope}
  \DrawGrid
  \HLrow{2}{teal}\HLrow{3}{teal}\HLrow{5}{teal}
  \NZ{1/1, 1/5, 5/2, 5/6, 2/3, 2/1, 3/5, 3/3, 6/3}
\end{scope}
\draw[->,thick] (6.5,3) -- (7.5,3);
\begin{scope}[xshift=8cm]
  \DrawGrid
  \HLrow{2}{teal}\HLrow{3}{teal}\HLrow{5}{teal}
  \NZ{1/1, 1/5, 2/2, 2/6, 3/3, 3/1, 5/5, 5/3, 6/3}
\end{scope}
\end{tikzpicture}
\caption{3OPT: exchange three rows: (6$\rightarrow$4)}
\end{subfigure}

\caption{Illustration of the two row/column-move types used in the iterative improvement stage. Each panel shows \emph{before} (left) and \emph{after} (right) on a $6\times 6$ dummy sparsity pattern; colored bands indicate the rows/blocks affected by the move. The values in the parentheses are the numbers of non-empty diagonals before and after the moves.}
\label{fig:compact-moves}
\end{figure}

\subsubsection{Efficient moves: Gains, losses, and candidate queues}
Recomputing the full diagonal histogram after every tentative move would be too
expensive. Instead, we evaluate each move incrementally. Consider a row or column
$u$ placed at position $p$ and tentatively moved to position $q$. We define
\begin{small}
\begin{align}
\textsc{Leave}(u,p)\;&=\;\#\{k:\texttt{d\_nnz}[k] \text{ from 1 to 0}\},\\
\textsc{Arrive}(u,q)\;&=\;\#\{k:\texttt{d\_nnz}[k] \text{ from 0 to 1}\}.
\end{align}
\end{small}
Intuitively, $\textsc{Leave}(u,p)$ counts the diagonals that disappear because of the move,
whereas $\textsc{Arrive}(u,q)$ counts the new diagonals created by the move. For a row move, let row $u$ be relocated from $p$ to $q$. For each nonzero
$a_{u,j}$, its current diagonal index is
$
k_{\mathrm{old}} = (\pi_C(j)-p)\bmod n,
$
and after the move, it would lie on
$
k_{\mathrm{new}} = (\pi_C(j)-q)\bmod n.
$
This nonzero contributes to {\sc{Leave}}$(u,p)$ only when $\texttt{d\_nnz}[k_{\mathrm{old}}]=1$,
since in that case, removing row $u$ empties that diagonal. Likewise, it contributes
to {\sc{Arrive}}$(u,q)$ only when $\texttt{d\_nnz}[k_{\mathrm{new}}]=0$, since in that case, placing
row $u$ at $q$ creates a previously empty diagonal.
Column moves are handled symmetrically. If column $u$ is moved from $p$ to $q$,
then for each nonzero $a_{i,u}$ we use the transposed structure and compute
\begin{align}
k_{\mathrm{old}} = (p-\pi_R(i))\bmod n,\\
k_{\mathrm{new}} = (q-\pi_R(i))\bmod n.
\end{align}
The move contributes to {\sc{Leave}}$(u,p)$ if
$\texttt{d\_nnz}[k_{\mathrm{old}}]$ equals to 1, and to {\sc{Arrive}}$(u,q)$ if
$\texttt{d\_nnz}[k_{\mathrm{new}}]$ equals to 0.

Using these primitives, the net effect of a tentative move is evaluated as
\[
\text{gain} \;=\; \sum \textsc{Leave} \;-\; \sum \textsc{Arrive},
\]
summed across all rows/columns (two or three) participating in the move. Hence, we only inspect the nonzeros of the moved rows or columns rather than recomputing the entire objective
from scratch. During probing, the corresponding updates to \texttt{d\_nnz} are applied
temporarily and rolled back immediately if the move is rejected.
\\

{\noindent{\bf {Focusing the search (candidate queues):}}} The efficiency of the implementation comes from moving only rows/columns that \emph{touch the scarcest
diagonals}. Let $\texttt{min\_nnz}$ be the current minimum positive occupancy in a non-empty diagonal.
We form two candidate queues $\mathcal{Q}_R,\mathcal{Q}_C$ by collecting rows
and columns incident to any diagonal $k$ with
$\texttt{d\_nnz}[k]\le \texttt{min\_nnz}+\beta$, using a small slack
$\beta$ (e.g., $\beta\in\{2,3,4\}$). This concentrates the search on entries
most likely to shrink $\#\text{diags}$ quickly. Within a sweep, we mark moved
indices to avoid reusing them until the next pass.

The process alternates \emph{column}/\emph{row} passes and iterates for a
small, fixed number of global sweeps. It stops early when $\#\text{diags} = \texttt{max\_deg}$ or if no move is accepted in a single pass. We now describe the mechanics of a move. 
\\

{\noindent{\bf 2OPT} ({\em from the column side - row is similar}):} 
Select $c_1\in\mathcal{Q}_C$ with current position $p_1$. Compute 
$g_1=\textsc{Leave}(c_1,p_1)$. If $g_1>0$ (or $g_1=0$ but the
secondary indicators do not worsen), scan a partner $c_2\neq c_1$ with position
$p_2$. For each potential partner, compute $g_2=\textsc{Leave}(c_2,p_2)$ where
\begin{small}
\begin{align*}
&\ell_{c_1\to p_2}=\textsc{Arrive}(c_1,p_2),\ 
\ell_{c_2\to p_1}=\textsc{Arrive}(c_2,p_1).
\end{align*}
\end{small}
The net change for exchanging the columns at positions $(p_1,p_2)$ is
\[
\Delta \;=\; g_1+g_2-\ell_{c_1\to p_2}-\ell_{c_2\to p_1}.
\]
We accept the move if $\Delta>0$, or $\Delta=0$ with respect to the primary and secondary metrics. We also mark these columns as moved and do not move them again in this local pass. Otherwise, the algorithm rejects the move and continues.\\

{\noindent{\bf 3OPT} ({\em from the column side - row is similar}):} 
\label{sec:three-opt}
When 2OPT stalls, a three–cycle can bypass the plateau. This is a common approach for optimization problems on graphs and matrices, such as the Traveling Salesman Problem~\citep{HELSGAUN2000106}. For $\pi_C$, pick an
ordered pair $(c_1,c_2)$ from $\mathcal{Q}_C$ with positions $(p_1,p_2)$ and
accumulated leave gain
$g_{12}=\textsc{Leave}(c_1,p_1)+\textsc{Leave}(c_2,p_2)$. Probe a third column
$c_3$ at $p_3$ (unmarked, i.e., unmoved in this pass). Consider the cyclic relocation
\[
c_1\!:\;p_1\to p_2,\qquad c_2\!:\;p_2\to p_3,\qquad c_3\!:\;p_3\to p_1,
\]
with $\textsc{Leave}(c_3,p_3)$ and arrival losses
$\ell_{c_1\to p_2},\ell_{c_2\to p_3},\ell_{c_3\to p_1}$. The net change is
\begin{small}
\[
\Delta \;=\; g_{12}+\textsc{Leave}(c_3,p_3)
           \;-\; \bigl(\ell_{c_1\to p_2}+\ell_{c_2\to p_3}+\ell_{c_3\to p_1}\bigr).
\]
\end{small}
3OPT aims to expand the neighborhood just enough to escape local minima created by pairwise exchanges. The algorithm accepts a move via the same primary/secondary rule as in 2OPT. Otherwise, the move is reverted. As in the description of 2OPT, the row version, which updates $\pi_R$, is symmetric and omitted here.

\begin{algorithm}[t]
\small
\caption{2DPP Iterative-Improvement}
\label{alg:global-iter-noflags}
\begin{algorithmic}[1]
\Require {The original matrix $\Ab$, \newline\hspace*{5ex}Initial permutations $(\pi_R,\pi_C)$, \newline\hspace*{5ex}Lower bound $\texttt{max\_deg}$, \newline\hspace*{5ex}Max number of global passes $\Gamma$, \newline\hspace*{5ex}Slack $\beta$; limits candidate queue sizes}
\Ensure {Final permutations $(\pi_R,\pi_C)$}
\newline
\State $(\texttt{d\_nnz},\#\texttt{diags},\texttt{min\_nnz},\texttt{min\_cnt})\gets\textsc{InitStats}(\Ab,\pi_R,\pi_C)$
\For{$y=1$ to $\Gamma$}
  \State $(\mathcal{Q}_R,\mathcal{Q}_C)\gets\textsc{ResetCand}(\texttt{d\_nnz},\texttt{min\_nnz},\beta)$
  \State \textsc{2OPT-Sweep}($\mathcal{Q}_R$); \textsc{2OPT-Sweep}($\mathcal{Q}_C$);
  \State $(\mathcal{Q}_R,\mathcal{Q}_C)\gets\textsc{ResetCand}(\texttt{d\_nnz},\texttt{min\_nnz},\beta)$
    \State \textsc{3OPT-Sweep}($\mathcal{Q}_R$); \textsc{3OPT-Sweep}($\mathcal{Q}_C$);
   \If {\#\texttt{diags} = \texttt{max\_deg}} 
   \textbf{break}
   \EndIf
   \If {No moves are accepted in this pass}
     \textbf{break}
   \EndIf
\EndFor
\State \Return $(\pi_R,\pi_C)$
\end{algorithmic}
\end{algorithm}

\subsubsection{The overall heuristic for 2DPP}
Algorithm~\ref{alg:global-iter-noflags} shows the high-level description of the overall process and how the parts described above are integrated with each other. The heuristic maintains $(\pi_R,\pi_C)$ together with a histogram $\texttt{d\_nnz}$ of nonzeros per cyclic diagonal. After initializing these statistics, it executes up to $\Gamma$ global passes. In each pass, \textsc{ResetCand} gathers row/column queues $(\mathcal{Q}_R,\mathcal{Q}_C)$ that touch the sparsest diagonals (within slack $\beta$). The algorithm then performs greedy 2OPT sweeps on columns and rows in these queues. The \textsc{Leave}/\textsc{Arrive} logic efficiently keeps track of the reduction in the number of occupied diagonals (ties are decided by improving $(\texttt{min\_nnz},\texttt{min\_cnt})$). After refreshing the queues, it runs 3OPT in a similar fashion. 
The process stops early (1) once $\#\texttt{diags}$ becomes equal to $\texttt{max\_deg}$, the maximum number of nonzeros within a row/column in the matrix, or (2) if no moves are accepted in a single pass. 
Otherwise, all $\Gamma$ passes are completed, and $(\pi_R,\pi_C)$ are returned.

\subsubsection{Complexity of the optimization}
Each attempted move touches only the nonzeros of the candidate rows/columns. Thanks to
the \textsc{Leave}/\textsc{Arrive} accounting, evaluation is $O\left(\sum_{i \in \mathcal{R} \cup \mathcal{C}} \nnz(i)\right)$, where $\mathcal{R}$ and $\mathcal{C}$ are the sets of rows and columns probed.
Furthermore, restricting the candidates to $\mathcal{Q}_R,\mathcal{Q}_C$ keeps the number of
attempts small. Overall, the expected runtime is linear in the number of probed rows and columns times the average degree. 

\section{Removing Nonzeros on the Dense Rows/Columns}\label{sec:dense}
A common bottleneck we have observed in our experiments is the presence of \emph{dense rows and/or columns}: even when most of the matrix is amenable, a single or a few densely populated rows/columns can cap the attainable optimization. Since the maximum row/column degree lower-bounds our objective~(see Eq.~\ref{eq:bound}), isolating these dense structures and handling them separately allows the optimizer to act effectively on the remaining sparse(r) core. In one drastic example, {\tt Chebyshev1} with $n = 261$ rows/columns and $\tau = 2319$ nonzeros\footnote{\url{https://sparse.tamu.edu/Muite}}, four rows are completely full. Hence, without any nonzero removal, the proposed optimizer stalls at $n$ diagonals. With the removal of full rows' nonzeros, processing the sparse core and then adding back the four row dot-products reduces the number of diagonals to $5$. On our system, multiplying $5$ diagonals in encrypted form takes $\approx 120$\,ms (instead of processing $261$ diagonals, which takes $\approx 6250$\,ms). Additionally, processing four encrypted dense rows adds $\approx 200$\,ms to the runtime. Similarly, on the largest matrix from the same family, {\tt Chebyshev4} with $n = 68,121$ and $\tau = 5,377,761$ nonzeros, the process yields $n = 68,121$ diagonals when no nonzeros are removed. This number reduces to $2590$  after the nonzeros of the densest 15 rows are removed. The removed rows can then be independently processed to complete the SpMV operation.

The elimination can be formally defined as follows: Let $\mathbf{A}\in\mathbb{R}^{n\times n}$ and let $D_r\subseteq\{1,\ldots,n\}$ (dense rows) and $D_c\subseteq\{1,\ldots,n\}$ (dense columns) be indices selected by a heuristic. Let ${\mathbf{A}}'$ be the \emph{dissected} matrix obtained by erasing the nonzeros on these rows and columns. It contains the nonzeros:
\[
{a}_{ij}' \;=\;
\begin{cases}
0, & i\in D_r \;\text{or}\; j\in D_c,\\[2pt]
a_{ij}, & \text{otherwise}.
\end{cases}
\]
Given an input vector $\mathbf{x}$, one can first compute the core product 
 ${\mathbf{b}'} \;=\; \mathbf{A}'\,\mathbf{x}$
homomorphically. Then, for each dense row $i\in D_r$, compute the $i$th entry of the output dot product $b_i = \mathbf{A}_{i,:}\mathbf{x}$. Similarly, for each dense column $j\in D_c$, add the columnwise terms $\mathbf{b} \;\leftarrow\; \mathbf{b} \;+\; \mathbf{A}_{:,j}\, x_j$.
Finally, assemble
\[
b_i \;=\;
\begin{cases}
\langle \mathbf{A}_{i,:},\,\mathbf{x}\rangle, & i\in D_r,\\[4pt]
b_i' \;+\; \displaystyle\sum_{j\in D_c} A_{ij} x_j, & i\notin D_r.
\end{cases}
\]

\begin{table}[htbp]
\small
\centering
\begin{tabular}{l|r}
\hline
& \textbf{Average} \\
\textbf{Operation} & \textbf{Time (µs)} \\ \hline
Cipher addition & 119.4 \\
Plain–cipher mult. & 203.1 \\
Cipher–cipher mult. (total) & $T_{cmult}$ = 3814.3 \\
\quad Multiplication & 455.0 \\
\quad Relinearization & 2639.5 \\
\quad Rescale & 719.9 \\
Ciphertext rotation (average) & $T_{rot}$ = 11073.3 \\
\quad Max rotation time & 22144.0 \\
\quad Min rotation time & 2691.0 \\ \hline
\end{tabular}
\caption{Performance of CKKS operations with ring dimension 8192 and slot count 4096.}
\label{tab:benchmarktable}
\end{table}

When selecting dense rows/columns to be eliminated, we have to account for the overhead incurred by treating them separately and compare with the speedup they yield. To this end, we count the ciphertext--ciphertext multiplications and rotations (plaintext--ciphertext ops and additions are cheap). Let $T_{\mathrm{cmult}}$ and $T_{\mathrm{rot}}$ denote their average times; on our platform $T_{\mathrm{rot}}\approx 3\,T_{\mathrm{cmult}}$ (see Table~\ref{tab:benchmarktable}). The overheads due to processing a single, additional dense row/column are: 
\begin{itemize}
  \item \textbf{Dense row $i\in D_r$} (inner product): cost $\approx T_{\mathrm{cmult}}+\log_2(n) T_{\mathrm{rot}}$.
  \item \textbf{Dense column $j\in D_c$} (adding $\mathbf{A}_{:,j}x_j$):
      cost $\approx T_{\mathrm{cmult}}$ per column.
\end{itemize}
Hence, removing dense row/column nonzeros is only meaningful when
\begin{align*}
&\left(|D_r| \times \bigl(T_{\mathrm{cmult}}+\lceil\log_2(n)\rceil\,T_{\mathrm{rot}}\bigr)\right) + \\
&\left(|D_c| \times \,T_{\mathrm{cmult}}\right)
\;\le\;
\Delta_{elim} \times \bigl(T_{\mathrm{cmult}}+T_{\mathrm{rot}}\bigr)
\end{align*}
where $\Delta_{elim}$ is the difference between the number of non-empty diagonals before and after removing the nonzeros of the rows in $D_r$ and columns in $D_c$. Algorithm~\ref{alg:dense-elim} presents the greedy approach used to choose dense components. 

\begin{algorithm}[t]
\caption{Dense row/column elimination}
\label{alg:dense-elim}
\begin{algorithmic}[1]
\Require $\Ab$, the proposed ordering + optimization $\mathcal{H}$, 
$T_{\mathrm{cmult}},T_{\mathrm{rot}}$, candidate budget $K$
\Ensure Dense sets $D_r,D_c$
\State $D_r,D_c \gets \emptyset$
\State Run $\mathcal{H}$ on $A$ and estimate baseline cost $C_{\mathrm{best}}$
\State Sort rows/columns in decreasing degree order
\For{$k=1,\ldots,K$}
    \State Let $(D_r^{(k)},D_c^{(k)})$ be the first $k$ candidates
    \State Construct $\Ab^{(k)}$ by zeroing $D_r^{(k)}$ and $D_c^{(k)}$
    \State Run $\mathcal{H}$ on $\Ab^{(k)}$ and obtain diag. count $s_k$
    \State Estimate total cost
    \begin{align*}
    C_k = &s_k \times (T_{\mathrm{cmult}}+T_{\mathrm{rot}})\ +\\ 
    &|D_r^{(k)}| \times  (T_{\mathrm{cmult}}+\lceil \log_2(n)\rceil T_{\mathrm{rot}})\ +\ \\ 
    &|D_c^{(k)}| \times T_{\mathrm{cmult}}.
    \end{align*}
    \If{$C_k < C_{\mathrm{best}}$}
    \vspace*{0.5ex}
        \State Store $(D_r^{(k)},D_c^{(k)})$ and update $C_{\mathrm{best}}$
    \EndIf
\EndFor
\State \Return best stored $(D_r,D_c)$
\end{algorithmic}
\end{algorithm}

\section{Experimental Results}\label{sec:exp}

To demonstrate the benefits of the proposed ordering, optimization, and row/column elimination techniques, we performed various experiments. Overall, we will focus on the following claims and conclusions; {\bf(1)} good initial orderings matter, {\bf(2)} local search is necessary, {\bf(3)} number of diagonals is a valid metric for the encrypted SpMV cost, and {\bf(4)} dense elimination addresses a separate performance obstruction. \\

\noindent {\bf{\underline{Dataset}}}: We used all square matrices with $10,000 \leq n \leq 50,000$ and $3 \leq \sigma \leq 20$ from SuiteSparse Matrix Collection~\citep{suitesparse}, where $\sigma = \frac{1}{n}\sum_{i=1}^{n}\delta_R(i) = \frac{1}{n}\sum_{j=1}^{n}\delta_C(j)$ is the average number of nonzeros per row/column in the matrix, $\delta_R(i) = |\{j: a_{i,j}\neq 0\}|$ and $\delta_C(j) = |\{i: a_{i,j}\neq 0\}|$ are the {\em number of nonzeros} of row $i$ and column $j$, respectively. To ensure that the optimizer has meaningful opportunities for improvement, we filter the dataset to retain only matrices whose number of non-empty diagonals in its {\em Natural} form is larger than $4 \times \max\left\{\max_{0 \leq i < n}{\delta_R(i)},\max_{0 \leq j < n}{\delta_C(j)}\right\}$  Overall, the dataset contains 175 matrices having an average of 8.7 per row/column. \\

\noindent {\bf{\underline{Experimental platform}}}:
All experiments were conducted on a server equipped with two 56-core \texttt{Intel(R) Xeon(R) Platinum 8480+} 2.00\,GHz CPUs, providing 112 HW threads with Hyper-Threading, 105\,MB last-level cache, and 256\,GB DDR5 memory. On this architecture, the symmetrization and the initial ordering are relatively cheap compared to the reduction in the encrypted {SpMV} time. On the other hand, for a practical experimental setting, the optimization timeout is limited to 1 hour. 

\subsection{Evaluating Relative Performance of the Variants}

Table~\ref{tab:global-variant-leaderboard} presents the results of the experiments with all 175 matrices, both symmetrization techniques, and all the orderings~(except {\sc{EigenOrder}}, which will be analyzed later). Overall, there are 24 variants in the table; 21 of them are {\em singleton}, i.e., using only a single symmetrization and initial ordering pair, whereas the remaining three, denoted by {\bf{*}} are {\em combined} variants that apply all the singletons and keep the best ordering found in terms of the number of diagonals. 

\begin{table*}[htbp]
\centering
\small
\scalebox{0.92}{
\begin{tabular}{ll|r||rr|r||rr||rr}
\toprule
& & & \multicolumn{3}{c||}{Leaderboard} & \multicolumn{2}{c|}{Perf vs.} & \multicolumn{2}{c}{Perf vs.} \\
\cmidrule(lr){4-6} 
\multirow{2}{*}{Sym.} & \multirow{2}{*}{Ord.} & \multirow{2}{*}{Opt.} & \multicolumn{2}{c|}{Win} & Avg. & \multicolumn{2}{c|}{Natural+NoOPT} & \multicolumn{2}{c}{Natural+3OPT} \\
\cmidrule(lr){7-8} \cmidrule(lr){9-10}
& & & {Count} & \multicolumn{1}{c|}{Perc.} & {Rank} & {AMean} & {Max} & {AMean} & {Max} \\
\midrule
\multicolumn{2}{c|}{*} & 3OPT & 160 & 91.43 & 1.17  & 5.50 & 45.61 & 3.41 & 28.29 \\
\multicolumn{2}{c|}{*}  & 2OPT & 58  & 33.14 & 2.67  & 5.47 & 45.61 & 3.39 & 28.29 \\
Pat.  & MP  & 3OPT  & 49 & 28.00 & 8.48  & 4.52 & 44.40 & 2.79 & 27.54 \\
Bpar. & RCM & 3OPT  & 38 & 21.71 & 7.58  & 4.88 & 45.61 & 3.01 & 28.29 \\
\multicolumn{2}{c|}{Natural} & 3OPT  & 31 & 17.71 & 13.94 & 1.49 & 3.50  & 1.00 & 1.00 \\
\multicolumn{2}{c|}{*} & NoOPT & 30 & 17.14 & 8.06  & 5.08 & 45.61 & 3.14 & 28.29 \\
Bpar. & LBS & 3OPT  & 20 & 11.43 & 8.23  & 4.82 & 44.52 & 2.98 & 27.61 \\
Bpar. & MP  & 3OPT  & 20 & 11.43 & 8.23  & 4.87 & 45.48 & 3.01 & 28.21 \\
Pat.  & RCM & 3OPT  & 18 & 10.29 & 10.23 & 4.39 & 45.61 & 2.75 & 28.29 \\
\multicolumn{2}{c|}{Natural} & 2OPT  & 17 & 9.71  & 14.81 & 1.45 & 3.46  & 0.97 & 1.07 \\
Bpar. & RCM & 2OPT  & 16 & 9.14  & 9.37  & 4.84 & 45.61 & 2.99 & 28.29 \\
Bpar. & MP  & 2OPT  & 13 & 7.43  & 9.83  & 4.84 & 45.48 & 2.99 & 28.21 \\
Bpar. & RCM & NoOPT & 11 & 6.29  & 13.79 & 4.68 & 45.61 & 2.88 & 28.29 \\
Pat.  & MP  & 2OPT  & 10 & 5.71  & 10.18 & 4.47 & 44.40 & 2.76 & 27.54 \\
Bpar. & LBS & 2OPT  & 9  & 5.14  & 9.84  & 4.79 & 44.58 & 2.96 & 27.65 \\
Pat.  & RCM & 2OPT  & 9  & 5.14  & 11.53 & 4.36 & 45.61 & 2.74 & 28.29 \\
\multicolumn{2}{c|}{Natural} & NoOPT & 8  & 4.57  & 18.84 & 1.00 & 1.00  & 0.71 & 1.00 \\
Bpar. & MP  & NoOPT & 7  & 4.00  & 14.36 & 4.68 & 45.48 & 2.88 & 28.21 \\
Pat.  & RCM & NoOPT & 7  & 4.00  & 15.86 & 4.13 & 45.61 & 2.59 & 28.29 \\
Bpar. & LBS & NoOPT & 5  & 2.86  & 14.38 & 4.62 & 44.40 & 2.85 & 27.54 \\
Pat.  & LBS & 3OPT  & 5  & 2.86  & 17.64 & 1.09 & 3.32  & 0.72 & 2.82 \\
Pat.  & MP  & NoOPT & 3  & 1.71  & 16.51 & 3.90 & 39.54 & 2.40 & 24.53 \\
Pat.  & LBS & 2OPT  & 1  & 0.57  & 18.83 & 1.06 & 3.52  & 0.70 & 3.07 \\
Pat.  & LBS & NoOPT & 0  & 0.00  & 22.13 & 0.76 & 2.30  & 0.52 & 1.97 \\
\bottomrule
\end{tabular}}
\caption{\small{The table summarizes the global performance of each of the 24 variants~(columns 1--3) across all matrices based on the number of diagonals, where a lowest diagonal count implies the first rank, i.e., a win~(columns 4--5, higher is better). The average rank is given in column 6~(lower is better). The first column, {\em Sym.}, identifies the symmetrization; 
{\em{Pat.}} is $\mathbf{\hat{B}} = \mathbf{B} + \mathbf{B}^\top$, and {\em{Bpar.}} is $\mathbf{\hat{B}} = \begin{bmatrix}
\mathbf{0} & \mathbf{B} \\
\mathbf{B}^\top & \mathbf{0}
\end{bmatrix}$.
The second column, {\em Ord.}, is the initial ordering. For these columns, {\em Natural} is the variant using the original ordering/permutation, and * is the one using the best symmetrization-initial ordering pair w.r.t. the diagonal count. The third column, {\em Opt.}, is the optimization level~(NoOPT, 2OPT, or 3OPT) applied later. Columns 8 and 9, respectively, present the average and maximum performance by normalizing the diagonal count of the baseline with no ordering or optimization to that of the variant~(Natural+NoOPT). The last two columns do the same with respect to the baseline Natural+3OPT.}}
\label{tab:global-variant-leaderboard}
\end{table*}

Table~\ref{tab:global-variant-leaderboard} shows that the variant *+NoOPT is able to win {\em only} in 30/175 matrices, hence, optimization is necessary; as expected, 3OPT is the most effective setting. Furthermore, the ranking of the three optimization levels is monotone; for instance, when the Natural ordering is used as the initial one, using optimization improves the win-counts from 8 (NoOPT) to 17 (2OPT) to 31 (3OPT). The same pattern is observed in the two best {\em singleton} symmetrization/ordering variants; Pattern-MP, from 3 to 10 to 49, and Bpar.-RCM, from 11 to 16 to 38, respectively. Similarly, for the {\em combined} variants, i.e., the $*$ rows, the win counts become 30, 58, and 160, respectively, where the last one corresponds to the 91.4$\%$ of the matrices. 

\begin{figure*}[htbp]
    \centering
    \includegraphics[width=0.8\textwidth]{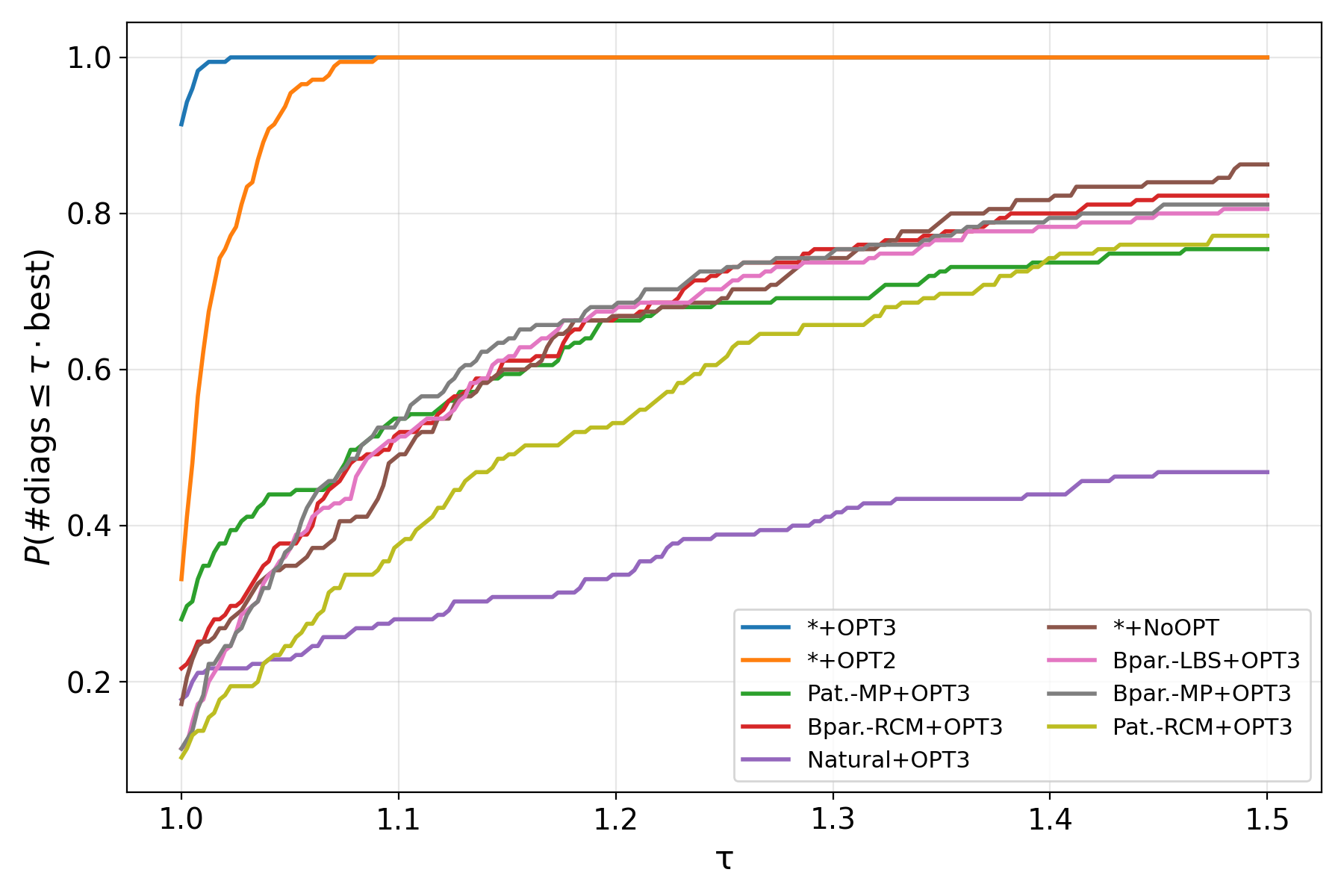}
    \caption{Performance profiles of the top nine variants from Table~\ref{tab:global-variant-leaderboard}. For each value of $\tau$, the y-axis shows the fraction of matrices for which a variant achieves a diagonal count at most $\tau$ times the best diagonal count obtained on that matrix.}
    \label{fig:perfprof}
\end{figure*}

Although the table shows that optimization is indispensable, with 2OPT already providing clear benefits and 3OPT delivering the greatest overall improvements, it also shows that optimization alone is not sufficient. The best variant without a proposed initial (re)ordering, Natural+3OPT, remains below the best reordered/optimized cases. In particular, the last two columns, which show the performance over Natural+3OPT for each variant, state that Bpar.-RCM+3OPT yields, on average, 3.01$\times$ fewer diagonals than
Natural+3OPT, and up to 28.29$\times$ fewer on a single matrix. Even Bpar.-RCM+NoOPT, which only takes the Bpar.-RCM output into account and does not perform any optimization, produces $2.88\times$ fewer diagonals on average compared to Natural+3OPT. These comparisons show that even when the strongest optimization strategy is applied, the initial permutation has a decisive impact on the performance. In other words, optimization is necessary, but it does not eliminate the importance of ordering; rather, it amplifies the advantage of good initial orderings.  For a more thorough analysis and to strengthen the observations above, Figure~\ref{fig:perfprof} shows the performance profiles of the top 9 variants in the table. The highest performance is obtained only when both ingredients are present together, namely, a strong optimization scheme and a suitable ordering strategy, or a strategy that tries all orderings. 

Table~\ref{tab:global-variant-leaderboard} suggests that Pat.-MP and Bpar.-RCM are the two most effective ordering choices, but for different reasons. Pat.-MP attains the best explicit reordered result under 3OPT, with 49 win counts and $2.79\times$ fewer diagonals compared to Natural+3OPT. It is an optimization-sensitive configuration: its win count rises from 10 under 2OPT to 49 under 3OPT, which means that 3OPT yields a $4.9\times$ improvement over 2OPT in terms of win count. This behavior suggests that Pat.-MP  has high potential realized when paired with strong optimization. By contrast, Bpar.-RCM only wins in 38 matrices; however, its relative performance w.r.t. Natural+3OPT, $3.01\times$, is slightly better than that of Pat.-MP. This is due to its {\em robustness}; the win count increases from 16 to only 38 when the optimization strategy is changed from 2OPT to 3OPT. All these make Pat.-MP and Bpar.-RCM rewarding singleton choices and probably sufficient for good performance.
Figure~\ref{fig:perfprof} also verifies these observations; when $\tau = 1.05$, Pat.-MP's diagonal count is less than or equal to $\tau \times best$ in $44.0\%$ of the matrices, where $best$ is the diagonal count of the best permutation for a matrix. On the other hand, when $\tau = 1.5$, among the singleton variants, Bpar.-RCM robustly leads the analysis with $82.3\%$ of the matrices, whereas Pat.-MP ranks only 8th with $77.1\%$.

Since the encrypted {SpMV} is an expensive kernel, usually performed multiple times, going over all the initial reorderings and keeping the best one, *+3OPT in our experiments, is a valid and superior choice. As we will state later, the execution time of the initial orderings is much cheaper compared to encrypted {SpMV} and optimization.

\subsubsection{Impact on the Encrypted SpMV}

Among the 175 matrices, the matrices with the top 20 highest gains on the encrypted SpMV time are given in Table~\ref{tab:diag_client_time_results}. The table shows that without performing any other technique, just by reordering the rows/columns, the SpMV time can be reduced by 25$\times$. In the table, the baseline is not a random permutation but the variant Natural+NoOPT. Hence, although the natural ordering can be superior from time to time (8 win counts in Table~\ref{tab:global-variant-leaderboard}), when it is not effective, there can be a significant performance boost thanks to the proposed techniques in the work. 

\begin{table*}[htbp]
\setlength{\tabcolsep}{4pt}
\centering
\scalebox{0.87}{
\begin{tabular}{l r r|l |r r |r r r}
\toprule
\multicolumn{3}{c|}{} & \multicolumn{1}{c|}{}  & \multicolumn{2}{c|}{\#diags} & \multicolumn{3}{c}{SpMV time~(sec)}  \\
 \multicolumn{1}{l}{Matrix} & \multicolumn{1}{c}{$n$} & \multicolumn{1}{c|}{$nnz$} & \multicolumn{1}{c|}{Ordering} & \multicolumn{1}{c}{$init$} & \multicolumn{1}{c|}{$best$} & \multicolumn{1}{c}{$init$} & \multicolumn{1}{c}{$best$} & Gain \\
\midrule
{\tt tuma1}                & 22967 &  87760 & Pat.-MP+3OPT & 17629 &  687 & 1574.7 &  61.4 & 96.1\% \\
{\tt delaunay\_n15}        & 32768 & 196548 & Pat.-MP+3OPT & 26035 & 1419 & 3100.8 & 169.0 & 94.5\% \\
{\tt de2010}               & 24115 & 116056 & Pat.-MP+3OPT & 17448 &  966 & 1558.6 &  86.3 & 94.5\% \\
{\tt ri2010}               & 25181 & 125750 & Pat.-RCM+3OPT & 20628 & 1330 & 2149.7 & 138.6 & 93.5\% \\
{\tt OPF\_10000}           & 43887 & 467711 & Pat.-MP+3OPT & 41965 & 2982 & 6872.3 & 488.3 & 92.9\% \\
{\tt delaunay\_n14}        & 16384 &  98244 & Pat.-MP+3OPT & 13235 &  956 &  788.2 &  56.9 & 92.8\% \\
{\tt hi2010}               & 25016 & 124126 & Pat.-MP+3OPT & 15362 & 1436 & 1600.9 & 149.7 & 90.7\% \\
{\tt viscoplastic2}        & 32769 & 381326 & Pat.-RCM+3OPT & 29689 & 2824 & 3978.0 & 378.4 & 90.5\% \\
{\tt worms20\_10NN}        & 20055 & 240826 & Bpar.-RCM+2OPT & 16520 & 1869 & 1229.7 & 139.1 & 88.7\% \\
{\tt powersim}             & 15838 &  67562 & Pat.-MP+3OPT &  9068 & 1079 &  540.0 &  64.3 & 88.1\% \\
{\tt descriptor\_xingo}  & 20738 &  73916 & Bpar.-RCM+3OPT & 12764 & 1687 & 1140.2 & 150.7 & 86.8\% \\
{\tt xingo3012}            & 20944 &  74386 & Bpar.-LBS+3OPT & 12744 & 1708 & 1138.4 & 152.6 & 86.6\% \\
{\tt bayer04}              & 20545 & 159082 & Bpar.-RCM+3OPT & 19863 & 2783 & 1774.3 & 248.6 & 86.0\% \\
{\tt hvdc1}                & 24842 & 159981 & Pat.-MP+3OPT & 19711 & 2809 & 2054.2 & 292.7 & 85.8\% \\
{\tt waveguide3D}          & 21036 & 303468 & Pat.-MP+3OPT & 21036 & 3049 & 1879.1 & 272.4 & 85.5\% \\
{\tt bcsstm36}             & 23052 & 320606 & Bpar.-MP+3OPT & 10458 & 1548 &  934.2 & 138.3 & 85.2\% \\
{\tt nopss\_11k}           & 11685 &  44941 & Bpar.-RCM+3OPT  &  6584 & 1038 &  294.1 &  46.4 & 84.2\% \\
{\tt bips07\_3078}         & 21128 &  75729 & Bpar.-RCM+3OPT  & 10717 & 1720 &  957.3 & 153.6 & 84.0\% \\
{\tt bips98\_1450}         & 11305 &  44678 & Bpar.-RCM+3OPT  &  6785 & 1117 &  303.0 &  49.9 & 83.5\% \\
{\tt bips07\_1693}         & 13275 &  49044 & Bpar.-RCM+3OPT  &  7004 & 1184 &  417.1 &  70.5 & 83.1\% \\
\bottomrule
\end{tabular}}
\caption{Baseline (no ordering) vs. best diagonal count and client time results.}
\label{tab:diag_client_time_results}
\end{table*}

\subsection{Impact of Dense Row/Column Removal}

\begin{figure*}[h]
    \centering
    \includegraphics[width=0.9\textwidth]{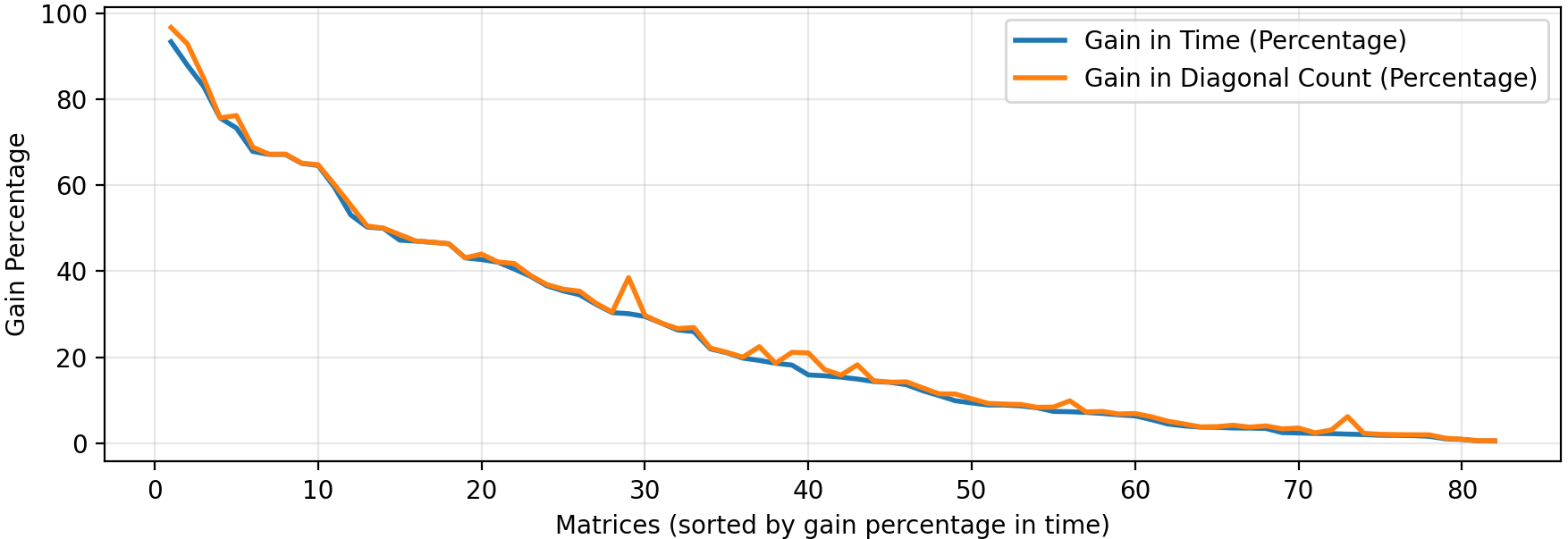}
    \caption{Estimated execution-time gain versus diagonal-count reduction for the 3OPT variant, with one point per matrix. Matrices are sorted by decreasing execution-time gain; the y-axis reports percentage improvement relative to the non-eliminated case. The figure shows that reductions in diagonal count generally track runtime improvement, but the relationship is not perfectly linear across matrices.}
    \label{fig:removal}
\end{figure*}

Among the 175 matrices in the experimental setting, row/column elimination has a positive impact on 84 matrices. Hence, the proposed elimination technique is broadly effective. Figure~\ref{fig:removal} shows the reduction in the execution time and the number of diagonals sorted with respect to the former. For these matrices, compared to the best variant without elimination, the average reduction on the encrypted SpMV time is 25.1\%, whereas the average reduction on the diagonal count is 26.0\%.
Overall, about one-third of matrices see $>30\%$ runtime reduction, while roughly a quarter experience $\le 5\%$ improvement, indicating elimination can be valuable when it causes substantial diagonal collapse. Table~\ref{tab:diag_reduction_comparison} presents the diagonal counts and execution times for the top 20 most impacted matrices. The matrices are sorted with respect to the reduction percentage compared to the best variant without elimination. Hence, the improvement is only due to the elimination. For instance, for the matrix {\tt memplus} with originally 16878 diagonals, the best variant without elimination provides around 48\% improvement on the encrypted SpMV time. However, with elimination, the SpMV operation becomes 23.7$\times$ faster. 

\begin{table*}[htbp]
\small
\centering
\caption{The impact of row/column elimination on the encrypted SpMV on 20 matrices with the highest gains; the first three columns describe the matrix. Columns 4-5 provide the number of diagonals and encrypted SpMV time for the natural ordering, whereas columns 6-7 do the same for the best variant without row/column elimination. The number of eliminated dense rows/columns and the remaining number of nonzeros are given in 8-9, where the next two columns provide the diagonal count and SpMV time, including the dense row/column processing. The last two columns provide the reduction percentages over the best variant without elimination.}
\label{tab:diag_reduction_comparison}
\setlength{\tabcolsep}{3pt}
\scalebox{0.90}{
\begin{tabular}{lrr||rr||rr||rr|rr||rr}
\toprule
        &       &           & \multicolumn{2}{c||}{Original} & \multicolumn{2}{c||}{Wout. Elim. } & \multicolumn{4}{c||}{With Row/Col Elim.} & \multicolumn{2}{c}{Reduction} \\
        &       &           & \multicolumn{2}{c||}{SpMV}  &  \multicolumn{2}{c||}{Best SpMV} & & &  \multicolumn{2}{c||}{SpMV} & \#diags & Time \\
Matrix  & $n$   & $nnz$     & \#diags & Time & \#diags & Time & \#elim & $nnz$ & \#diags & Time & \% & \% \\
\midrule
{\tt memplus}        & 17758 & 126150 & 16878 & 1256.4 &  8835 &  657.7 &   162 &  72924 &   357 &   53.1 & 96.0 & 91.9 \\
{\tt as-22july06}    & 22963 &  96872 & 22360 & 1997.3 & 12774 & 1141.1 &   363 &  21490 &   979 &  148.5 & 92.3 & 87.0 \\
{\tt as-caida}       & 31379 & 106762 & 30192 & 3595.9 & 15838 & 1886.3 &   226 &  32432 &  2627 &  352.0 & 83.4 & 81.3 \\
{\tt c-52}           & 23948 & 202716 & 23376 & 2088.1 & 15268 & 1363.8 &     1 & 199289 &  3735 &  333.8 & 75.5 & 75.5 \\
{\tt c-46}           & 14913 & 130397 & 13991 &  833.2 &  7649 &  455.5 &   100 &  57799 &  2221 &  148.4 & 71.0 & 67.4 \\
{\tt c-65}           & 48066 & 360528 & 46450 & 8298.4 & 26937 & 4812.3 &    12 & 346002 &  9299 & 1663.4 & 65.5 & 65.4 \\
{\tt c-54}           & 31793 & 391693 & 29211 & 3479.1 & 19633 & 2338.3 &    16 & 351853 &  6918 &  826.7 & 64.8 & 64.6 \\
{\tt ckt11752}       & 49702 & 333029 & 36238 & 7013.5 & 11577 & 2240.6 &    21 & 303564 &  4078 &  793.0 & 64.8 & 64.6 \\
{\tt rajat15}        & 37261 & 443573 & 36849 & 5485.9 & 22726 & 3383.4 &    11 & 415956 &  8921 & 1330.1 & 60.8 & 60.7 \\
{\tt c-53}           & 30235 & 372213 & 29055 & 3460.5 & 15081 & 1796.2 &   133 & 140018 &  6029 &  741.0 & 60.0 & 58.8 \\
{\tt c-44}           & 10728 &  85000 & 10403 &  464.6 &  6222 &  277.9 &    11 &  81335 &  2535 &  114.9 & 59.3 & 58.6 \\
{\tt c-49}           & 21132 & 157042 & 20251 & 1808.9 & 11566 & 1033.1 &     3 & 153197 &  5780 &  516.8 & 50.0 & 50.0 \\
{\tt c-66b}          & 49989 & 499007 & 48139 & 9316.8 & 31175 & 6033.6 &    84 & 410797 & 15734 & 3060.3 & 49.5 & 49.3 \\
{\tt c-60}           & 43640 & 298578 & 41912 & 6863.7 & 18168 & 2975.3 &     1 & 294163 &  9931 & 1626.5 & 45.3 & 45.3 \\
{\tt c-61}           & 43618 & 310016 & 40891 & 6696.5 & 22132 & 3624.4 &     1 & 304357 & 12101 & 1981.9 & 45.3 & 45.3 \\
{\tt c-50}           & 22401 & 193625 & 21323 & 1904.7 & 12105 & 1081.3 &     2 & 186769 &  7191 &  642.7 & 40.6 & 40.6 \\
{\tt c-42}           & 10471 & 110285 & 10237 &  457.2 &  5438 &  242.9 &    45 &  66786 &  3088 &  144.9 & 43.2 & 40.3 \\
{\tt c-63}           & 44234 & 434704 & 43488 & 7121.8 & 29559 & 4840.7 &    13 & 422837 & 17681 & 2897.8 & 40.2 & 40.1 \\
{\tt email-Enron}    & 36692 & 367662 & 36547 & 4896.9 & 25331 & 3394.1 &   263 & 212054 & 15592 & 2135.3 & 38.5 & 37.1 \\
{\tt c-55 }          & 32780 & 403450 & 32538 & 4359.7 & 22989 & 3080.3 &    49 & 379951 & 14579 & 1961.9 & 36.6 & 36.3 \\
\bottomrule
\end{tabular}
}
\end{table*}

\subsection{The Cost of the Ordering}

For all the ordering heuristics on the 175 matrices in our dataset, the mean execution time is less than 1 second, which is negligible compared to the optimization time with a timeout of 3600 seconds. For both optimization levels, Pat.-LBS and Natural orderings yield the most time for optimization; for 2OPT, the median execution times are 678 and 611 seconds, respectively, whereas for 3OPT, the execution times are 3600 and 3336 seconds, respectively. The next highest median optimization times across all the matrices are 212 and 1672 seconds for 2OPT and 3OPT, respectively. Pat.-LBS is the worst variant, placed in the last 3/4 rows of Table~\ref{tab:global-variant-leaderboard}. This suggests that, on Pat.-LBS, the optimization mostly explores unproductive moves. On the other hand, for Natural, the optimization seems to be more fruitful.

\section{Related Work}\label{sec:review}
The \cite{halevishoup} matrix-vector multiplication method was one of the first matrix operation optimizations in FHE schemes, which was later further optimized by an adaptation of the baby-step giant-step algorithm~\citep{HaleviShoupBSGS}. Its use of modular diagonals can be directly linked to combinatorial structures, such as circulant graphs studied by \cite{AntonyNaduvath2022CirculantCompletion, AntonyNaduvath2024CirculantCompletion}. Aside from Halevi-Shoup, many more studies have been made \citep{Rizomiliotis2022, Mahon2025, ChenBicyclic2024, Huang2023, ParkCCMM2025, Jiang2018} on HE-based matrix multiplication in recent years. These works primarily focus on dense matrix operations, utilizing techniques such as SIMD ciphertext packing, polynomial embedding, and block matrix decomposition to optimize the heavy rotational and multiplicative overhead. While effective for dense structures, a common limitation across these studies is that they do not account for the nonzero pattern. When applied to sparse matrices, these dense packing schemes perform unnecessary operations on zero elements.

The optimization of sparse matrix-vector multiplication (SpMV) has been extensively studied for plaintext execution on multicore CPUs and GPUs \citep{williams2007optimization, xu2010sparse}. In the plaintext domain, SpMV is fundamentally a memory-bound operation. Consequently, traditional optimization techniques rely on structural matrix reordering algorithms, such as RCM \citep{10.1145/800195.805928}, to cluster nonzeros around the main diagonal. The goal is to minimize memory bandwidth and maximize spatial cache locality. In stark contrast, memory hierarchy is not the primary bottleneck in homomorphically encrypted SpMV. Instead, the computational overhead is dominated by cipher-cipher multiplications and rotations.

The literature specifically addressing encrypted SpMV is much scarcer. Among the few existing works, \cite{Gao2025SpMVHE} and \cite{10.1145/3721146.3721948} explore compressing the nonzero elements of the sparse matrix directly into the ciphertexts to accelerate computations. While this approach eliminates redundant operations, it inherently leaks information about the topological structure of the matrix—or the underlying data it represents—because structural metadata must be transferred alongside the ciphertexts. Alternatively, \cite{Yu2025Lodia} proposed {\em Lodia}, which decomposes the sparse matrix into a series of low-diagonal matrices. This method successfully masks the exact sparsity pattern, preserving privacy, but the decomposition process exhausts significantly more ciphertext multiplication levels within the underlying HE scheme.

The optimization of encrypted SpMV is particularly critical for the deployment of privacy-preserving machine learning, where irregular sparsity forms a severe computational bottleneck. A primary example is the homomorphic evaluation of Graph Neural Networks (GNNs), where the core message-passing mechanism relies heavily on multiplying sparse adjacency matrices with dense feature vectors. Recent frameworks, such as CryptoGCN \citep{cryptogcn2022}, have demonstrated that the overhead caused by adjacency matrix multiplications can be mitigated by integrating matrix sparsity into the packing scheme. However, while these methods perform effectively on standard benchmark datasets, they struggle to scale to larger, more complex workloads. The highly irregular sparsity patterns found in massive, real-world applications—such as collaborative anti-money laundering (AML) networks—induce bottlenecks that outpace the capabilities of these localized packing strategies. Consequently, there remains a critical need for optimization techniques that can reliably handle the scale and dimensionality of industrial-grade graph datasets. 

\section{Conclusion and Future Work}\label{sec:conc}
In this work, we utilize sparsity to combat the computational bottleneck of encrypted SpMV. While traditional dense packing schemes redundantly process zeros, and existing sparse compression methods compromise structural privacy, here we propose a novel approach. By formalizing the 2-Dimensional Diagonal Packing Problem (2DPP), we fundamentally shift the objective of matrix reordering from achieving cache locality or load balancing to the minimization objective of Halevi-Shoup cyclic diagonals.

To solve 2DPP, we introduce a two-stage pipeline that adapts classical combinatorial graph algorithms, such as the Reverse Cuthill-McKee (RCM) and Miller-Pritikin (MP) ordering, into the encrypted domain. This initial reordering is then refined by our novel iterative-improvement heuristics. Our extensive empirical evaluations demonstrate that the *+3OPT variant dominates by obtaining the best diagonal count for 160/175 matrices in our experiments. For highly sparse matrices, our pipeline successfully collapsed the cyclic diagonal count by orders of magnitude (e.g., reducing nearly 10,000 initial diagonals down to the theoretical target bounds). Crucially, since our approach relies on cyclic diagonal packing rather than the actual nonzero structure, it does not reveal the exact sparsity pattern. 

At this stage, the techniques proposed are exclusively for square matrices. As a natural extension, future work will focus on adapting this method to rectangular matrices. One promising direction is to transform rectangular matrices into equivalent square forms through their bipartite graph representations. By applying our reordering strategy to these bipartite representations, we can derive corresponding row and column permutations for the original rectangular matrix. Overall, the 2DPP heuristics are designed to serve as a drop-in compiler optimization for frameworks reliant on encrypted SpMV, such as homomorphically evaluated Graph Neural Networks (GNNs).

\section{Appendix}\label{sec:appendix}

For this study, we aimed to increase the robustness of using {\sc EigenOrder} by utilizing the eigenvectors corresponding to the top $nev$ eigenvalues with Algorithm~\ref{alg:eigen-angle-order}. By sorting the rows and columns by the polar angle for each pair and selecting the permutation that yields the fewest resulting wraparound diagonals, we reliably recover the latent structure. 

\begin{figure}[htbp]
  \centering
  \begin{subfigure}[b]{0.48\linewidth}
    \includegraphics[width=\textwidth]{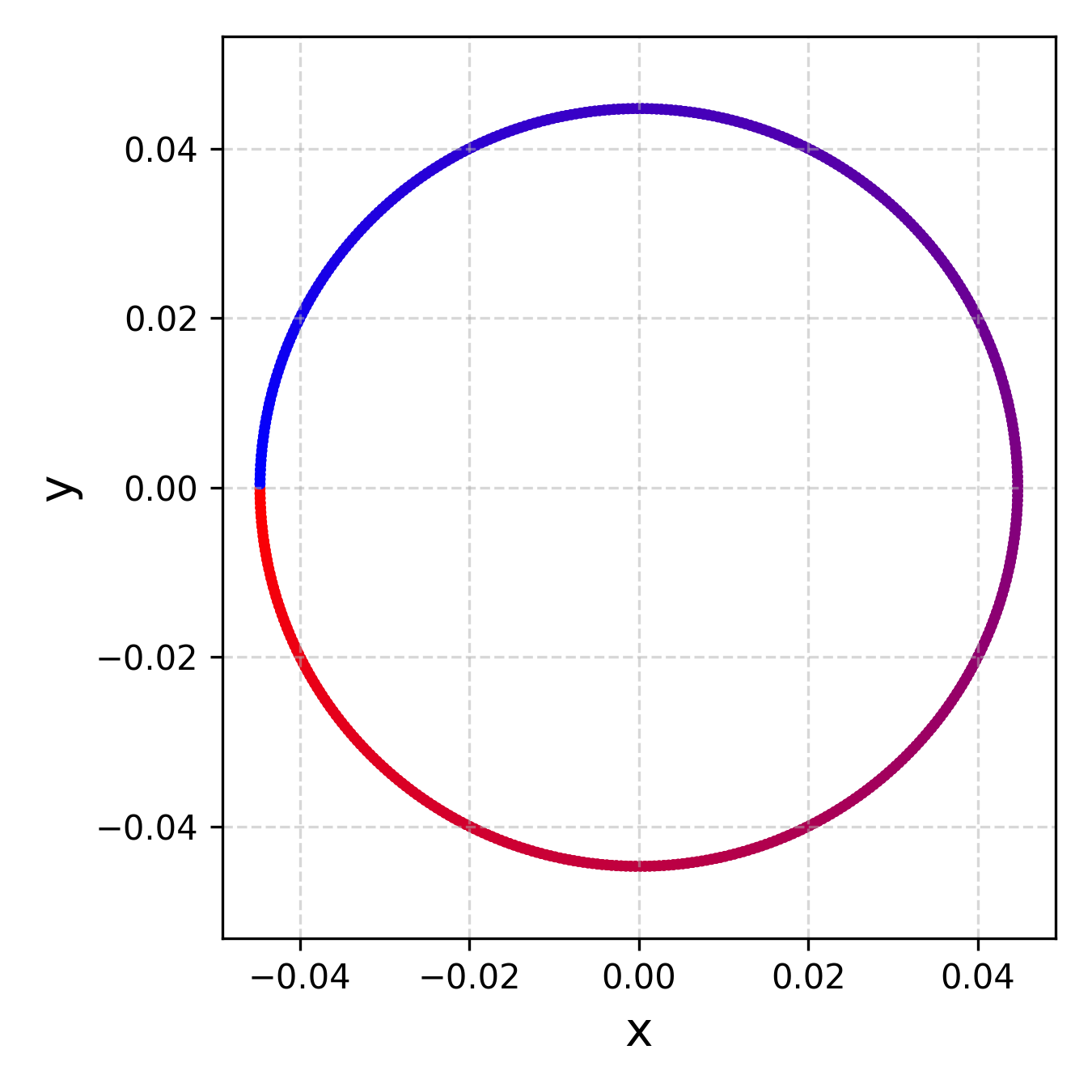}
    \caption{0\% noise}
    \label{fig:first}
  \end{subfigure}
  \begin{subfigure}[b]{0.48\linewidth}
    \includegraphics[width=\textwidth]{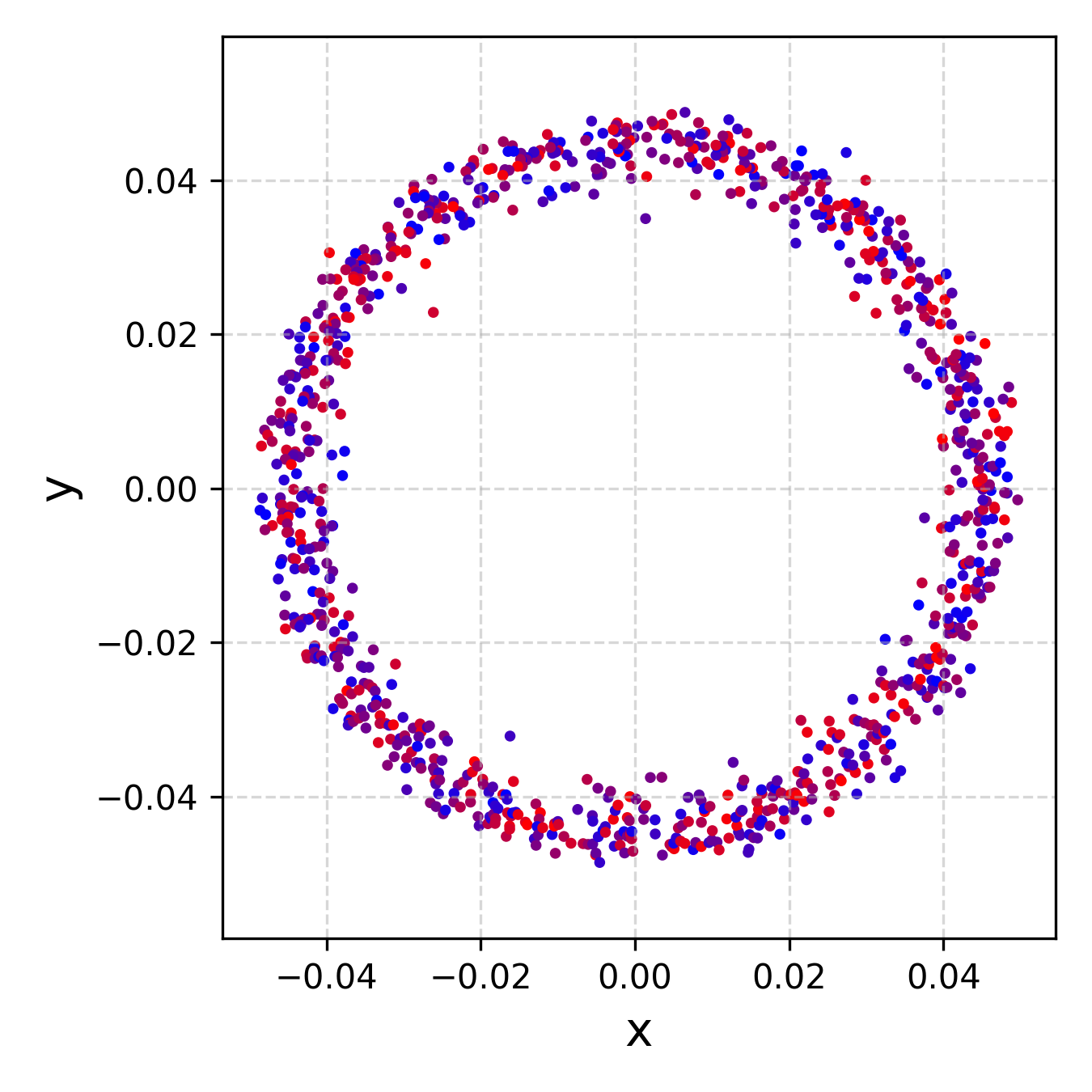}
    \caption{10\% noise}
    \label{fig:second}
  \end{subfigure}
  \caption{The plots of the $1000 \times 1000$ matrix with 50 diagonals and varying noise. For each row/column $i$, the coordinate $(x_i, y_i)$ is obtained from the corresponding eigenvector entries associated with the 2nd and 3rd largest eigenvalue. Each point is colored based on the order of the corresponding row/col in the original, symmetric, circulant matrix. The same color is used for the same row/col in the sub-figures.}
  \label{fig:plotssynth}
\end{figure}

\begin{algorithm}[htbp]
\small
\caption{$\pi^*_{eig} = $ {\sc EigenOrder}($\mathbf{\hat{B}}$, $nev$)}
\label{alg:eigen-angle-order}
\begin{algorithmic}[1]
\Require $\mathbf{\hat{B}}$, the symmetrized sparse matrix; $nev$, the number of top eigenvectors to evaluate
\Ensure Mapping $\pi^*_{eig}:V\!\to\!\{0,\dots,n-1\}$
\State Compute $E = \{\mathbf{u}_1, \dots, \mathbf{u}_{nev}\}$, the eigenvectors associated with the $nev$ largest algebraic eigenvalues of $\mathbf{\hat{B}}$
\State $min\_diags \gets \infty$
\For{each eigenvector pairs $(\mathbf{u}_i, \mathbf{u}_j)$ with $i < j$}
\State For each index $k \in V$, let $\varphi_k$ be the angular coordinate $\text{atan2}(\mathbf{u}_{j,k}, \mathbf{u}_{i,k})$
\State Define $\pi$ so that $\pi(k) > \pi(m)$ iff $\varphi_k \ge \varphi_m$
\State Apply perm. $\pi$ to $\mathbf{\hat{B}}$ to obtain $\mathbf{\hat{B}}_{\pi}$
\State $current\_diags \gets$ {\sc CountDiagonals}($\mathbf{\hat{B}}_{\pi}$)
\If{$current\_diags < min\_diags$}
\State $min\_diags \gets current\_diags$
\State $\pi^*_{eig} \gets \pi$
\EndIf
\EndFor
\State \Return $\pi^*_{eig}$
\end{algorithmic}
\end{algorithm}

As a preliminary experiment, we generate random, symmetric matrices ith only a few full diagonals to test the performance of {\sc EigenOrder}. To incur {\em noise}, we remove each edge of the implied graph with $p$ probability. We then randomly permute the rows and columns of this almost circulant matrix. A good algorithm for 2DPP should find the inverse of these permutations and revert the nonzeros to their original diagonals. 

In Figure~\ref{fig:plotssynth}, the points corresponding to the rows/columns of the circulant matrix are colored based on their initial order; the colors red and blue are used for the first and last rows/columns, respectively, in the original matrix. Hence, the same color is used for the same row/column in each of the three sub-figures. The perfect circle with $0\%$ noise and the scattering of the colors with positive noise shows that although {\sc EigenOrder} is a perfect algorithm with no noise, even with a $10\%$ alteration of the nonzeros, the initial ordering is perturbed drastically and is not easy to regenerate based on the angular information. 

\subsection{Experiments on {\sc{EigenOrder}}}

In our experiments, we found that {\sc EigenOrder} does not work well for the real-life matrices obtained from SuiteSparse and frequently yields diagonal counts much worse than the best, regardless of the optimization level used with it. To analyze its performance differently, as explained in Section~\ref{sec:eigen}, we generate a synthetic dataset by generating symmetric $n \times n$ circulant matrices $\Ab$ with full diagonals, and add noise by removing each nonzero with probability $p$ while keeping the matrix symmetric. The rows/columns of the matrix are then randomly permuted with a random permutation matrix $\Pb$, $\Ab' = \Pb\Ab\Pb^\top$, and {\sc EigenOrder} is applied to the pattern of $\Ab'$. 

Table~\ref{tab:final_diag_counts} summarizes the performance of {\sc EigenOrder} for $\ell = \{10, 50\}$ initial diagonals, $n = \{1000, 10000\}$, $p = \{0\%, 1\%, 2\%, 3\%, 4\%, 5\%\}$, and $\#ev = \{50, 200\}$ eigenvectors for a more robust performance~(see Sec.~\ref{sec:eigen}). As Table~\ref{tab:final_diag_counts} shows, in the noiseless setting, the algorithm fails to obtain the optimal number of diagonals only in the case, $n = 1000$, $\ell = 10$, and
$\#ev = 50$. Even in this case, it obtained a pattern with 22 diagonals, where none of the proposed ordering variants produce fewer than $4 \times \ell$ diagonals for these cases. However, with more noise, the number of diagonals obtained via {\sc EigenOrder} drastically increases. On the contrary, due to increased sparsity, the diagonal counts of RCM, MP, and LBS decrease as $p$ increases. For $\ell = 50$, when the $p = 1\%$ noise is added, the number of diagonals obtained by {\sc EigenOrder} increases by around $15\times$ and $150\times$, respectively, when $n = 1000$ and $10000$. In short, since the sparsity patterns of real-life matrices do not resemble those of circulant matrices, {\sc EigenOrder} does not perform well. 
 
\begin{table}[ht]
\centering
\caption{The diagonal counts obtained by the {\sc{EigenOrder}}; the {\em Init.} columns (cols. 3 and 6) denote the number of diagonals of $\Ab' = \Pb\Ab\Pb^\top$ where each $\Ab$ is an $n \times n$ synthetic matrix with $n \in \{1000, 10000\}$ generated with $\ell \in \{10, 50\}$ random diagonals and different noise levels ($p$). $\Pb$ is a random permutation matrix. The {\sc EigenOrder} algorithm is executed by using eigenvectors corresponding to the $\#ev$ eigenvalues of $\Ab'$ with the largest magnitudes.}
\label{tab:final_diag_counts}
\setlength{\tabcolsep}{3.8pt}
\begin{tabular}{cc|rrr|rrr}
\toprule
\multirow{2}{*}{$\ell$} & \multirow{2}{*}{$p$}& \multicolumn{3}{c|}{$n=1000$} & \multicolumn{3}{c}{$n=10000$} \\
\cmidrule(lr){3-5} \cmidrule(l){6-8}
 &   & Init. & $\#ev=$ & $\#ev=$ & Init. & $\#ev=$ & $\#ev=$ \\
  &   &  & $50$ & $200$ &  & $50$ & $200$ \\

\midrule
\multirow{6}{*}{10}
 & none  & 999  & 22  & 10  & 9997 & 10   & 10   \\
 & 1\% & 999  & 97  & 97  & 9997 & 1375  & 1375  \\
 & 2\% & 999  & 113 & 113 & 9997 & 1595 & 1595 \\
 & 3\% & 999  & 119 & 119 & 9997 & 1669 & 1669 \\
 & 4\% & 999  & 123 & 123 & 9997 & 1877 & 1877 \\
 & 5\% & 999  & 117 & 117 & 9997 & 1837 & 1837 \\
\midrule
\multirow{6}{*}{50}
 & none  & 999  & 50  & 50  & 9999 & 50   & 50   \\
 & 1\% & 999  & 738 & 738 & 9999 & 7552 & 7552 \\
 & 2\% & 999  & 820 & 820 & 9999 & 7906 & 7906 \\
 & 3\% & 999  & 833 & 833 & 9999 & 8142 & 8142 \\
 & 4\% & 999  & 868 & 868 & 9999 & 8416 & 8416 \\
 & 5\% & 999  & 886 & 886 & 9999 & 8608 & 8608 \\
\bottomrule
\end{tabular}
\end{table}

\section*{List of Abbreviations}
\begin{tabular}{ll}
HE    & Homomorphic Encryption \\
SpMV  & Sparse Matrix--Vector Multiplication \\
2DPP  & Two-Dimensional Diagonal Packing Problem \\
ILP   & Integer Linear Programming \\
SIMD  & Single Instruction, Multiple Data \\
RCM   & Reverse Cuthill--McKee \\
MP    & Miller--Pritikin \\
LBS   & Level-Based Sweep \\
BFS   & Breadth-First Search \\
2OPT  & Two-element swap local search \\
3OPT  & Three-element cyclic local search \\
CKKS  & Cheon--Kim--Kim--Song \\
\end{tabular}

\section*{Statements and Declarations}

\paragraph{Competing interests}
The authors declare that they have no competing interests.

\paragraph{Code and Data Availability}
The sparse matrices used in this study are from the SuiteSparse Matrix Collection.
The complete filtering criteria are given in the manuscript. The implementation used for the experiments is available from the corresponding
author upon request.

\paragraph{Author contributions}
Kemal Mutluergil and Deniz Elbek contributed to the implementation, experiments,
and manuscript preparation. Kamer Kaya and Erkay Savaş contributed to the problem
formulation, supervision, and manuscript revision. All authors reviewed and approved
the final manuscript.

\paragraph{Funding}
This research received no external funding.

\paragraph{Ethics approval and consent to participate} 
Not applicable.

\paragraph{Consent for publication} 
Not applicable.

\bibliography{main}

@InProceedings{halevishoup,
author="Halevi, Shai
and Shoup, Victor",
editor="Garay, Juan A.
and Gennaro, Rosario",
title="Algorithms in {HElib}",
booktitle="Advances in Cryptology -- CRYPTO 2014",
year="2014",
publisher="Springer Berlin Heidelberg",
address="Berlin, Heidelberg",
pages="554--571",
isbn="978-3-662-44371-2"
}

@inproceedings{HaleviShoupBSGS,
author = {Halevi, Shai and Shoup, Victor},
title = {Faster Homomorphic Linear Transformations in HElib},
year = {2018},
isbn = {978-3-319-96883-4},
publisher = {Springer-Verlag},
address = {Berlin, Heidelberg},
doi = {10.1007/978-3-319-96884-1_4},
booktitle = {Advances in Cryptology – CRYPTO 2018: 38th Annual International Cryptology Conference, Santa Barbara, CA, USA, August 19–23, 2018, Proceedings, Part I},
pages = {93–120},
numpages = {28},
location = {Santa Barbara, CA, USA}
}

@article{HeinrichHell1987Bandsize,
  author  = {K. Heinrich and P. Hell},
  title   = {On the problems of bandsize},
  journal = {Graphs and Combinatorics},
  year    = {1987},
  volume  = {3},
  pages   = {279--284},
  doi = {10.1007/BF01788550}
}

@article{Sudborough1986Bandsize,
  author  = {I. H. Sudborough},
  title   = {Complexity of bandsize},
  journal = {A.M.S. Abstracts},
  volume  = {7},
  year    = {1986},
  pages   = {15},
  note    = {Abstract \#825-05-618}
}

@incollection{ERDOS1988117,
title = {Bandwidth versus Bandsize},
editor = {Lars Dovling Andersen and Ivan Tafteberg Jakobsen and Carsten Thomassen and Bjarne Toft and Preben Dahl Vestergaard},
series = {Annals of Discrete Mathematics},
publisher = {Elsevier},
volume = {41},
pages = {117-129},
year = {1988},
booktitle = {Graph Theory in Memory of G.A. Dirac},
issn = {0167-5060},
doi = {10.1016/S0167-5060(08)70455-2},
author = {P. Erdös and P. Hell and P. Winkler}
}

@inproceedings{10.1145/800195.805928,
author = {Cuthill, E. and McKee, J.},
title = {Reducing the bandwidth of sparse symmetric matrices},
year = {1969},
isbn = {9781450374934},
publisher = {ACM},
address = {New York, NY, USA},
doi = {10.1145/800195.805928},
booktitle = {Proceedings of the 1969 24th National Conference},
pages = {157–172},
numpages = {16},
series = {ACM '69}
}

@article{AntonyNaduvath2024CirculantCompletion,
  author  = {Toby B. Antony and Sudev Naduvath},
  title   = {Further studies on circulant completion of graphs},
  journal = {Proyecciones Journal of Mathematics},
  volume  = {43},
  number  = {3},
  pages   = {761--773},
  year    = {2024},
  month   = jun,
  url     = {https://www.revistaproyecciones.cl/index.php/proyecciones/article/download/6034/4504/38716}
}

@article{leung1984some,
  title={On some variants of the bandwidth minimization problem},
  author={Leung, Joseph YT and Vornberger, Oliver and Witthoff, James D},
  journal={SIAM Journal on Computing},
  volume={13},
  number={3},
  pages={650--667},
  year={1984},
  publisher={SIAM},
  doi ={10.1137/0213040}
}

@article{raspaud2009antibandwidth,
  title={Antibandwidth and cyclic antibandwidth of meshes and hypercubes},
  author={Raspaud, Andr{\'e} and Schr{\"o}der, Heiko and S{\`y}kora, Ondrej and Torok, Lubomir and Vrt’o, Imrich},
  journal={Discrete Mathematics},
  volume={309},
  number={11},
  pages={3541--3552},
  year={2009},
  publisher={Elsevier},
  doi={10.1016/j.disc.2007.12.058}
}

@article{miller1989separation,
  title={On the separation number of a graph},
  author={Miller, Zevi and Pritikin, Dan},
  journal={Networks},
  volume={19},
  number={6},
  pages={651--666},
  year={1989},
  publisher={Wiley Online Library}
}

@article{https://doi.org/10.1002/nla.1859,
author = {Scott, Jennifer and Hu, Yifan},
title = {Level-based heuristics and hill climbing for the antibandwidth maximization problem},
journal = {Numerical Linear Algebra with Applications},
volume = {21},
number = {1},
pages = {51-67},
doi = {10.1002/nla.1859},
year = {2014}
}

@article{GONZAGADEOLIVEIRA2020106434,
title = {An ant colony hyperheuristic approach for matrix bandwidth reduction},
journal = {Applied Soft Computing},
volume = {94},
pages = {106434},
year = {2020},
issn = {1568-4946},
doi = {10.1016/j.asoc.2020.106434},
author = {S.L. {Gonzaga de Oliveira} and L.M. Silva}
}

@article{10.1137/S1064827500379215,
author = {Hager, William W.},
title = {Minimizing the Profile of a Symmetric Matrix},
year = {2002},
issue_date = {2002},
publisher = {Society for Industrial and Applied Mathematics},
address = {USA},
volume = {23},
number = {5},
issn = {1064-8275},
doi = {10.1137/S1064827500379215},
journal = {SIAM J. Sci. Comput.},
month = jan,
pages = {1799–1816},
numpages = {18}
}

@article{10.1145/592843.592844,
author = {Reid, John K. and Scott, Jennifer A.},
title = {Implementing Hager's exchange methods for matrix profile reduction},
year = {2002},
issue_date = {December 2002},
publisher = {ACM},
address = {New York, NY, USA},
volume = {28},
number = {4},
issn = {0098-3500},
doi = {10.1145/592843.592844},
journal = {ACM Trans. Math. Softw.},
month = dec,
pages = {377–391},
numpages = {15}
}

@article{HELSGAUN2000106,
title = {An effective implementation of the Lin–Kernighan traveling salesman heuristic},
journal = {European Journal of Operational Research},
volume = {126},
number = {1},
pages = {106-130},
year = {2000},
issn = {0377-2217},
doi = {10.1016/S0377-2217(99)00284-2},
author = {Keld Helsgaun}
}

@inproceedings{Yu2025Lodia,
  author    = {Jiping Yu and Kun Chen and Xiaoyu Fan and Yunyi Chen and Xiaowei Zhu and Wenguang Chen},
  title     = {Lodia: Towards Optimal Sparse Matrix-Vector Multiplication for Batched Fully Homomorphic Encryption},
  booktitle = {Proceedings of the 2025 ACM SIGSAC Conference on Computer and Communications Security (CCS '25)},
  year      = {2025},
  publisher = {ACM},
  address   = {New York, NY, USA},
  doi = {10.1145/3719027.376502}
}

@article{Gao2025SpMVHE,
title = {Efficient privacy-preserving sparse matrix-vector multiplication using homomorphic encryption},
journal = {Information Sciences},
volume = {739},
pages = {123180},
year = {2026},
issn = {0020-0255},
doi = {10.1016/j.ins.2026.123180},
author = {Yang Gao and Gang Quan and Wujie Wen and Scott Piersall and Qian Lou and Liqiang Wang},
}

@inproceedings{10.1145/3721146.3721948,
author = {Ferguson, Aidan and Gibson, Perry and D'Agata, Lara and McLeod, Parker and Yaman, Ferhat and Das, Amitabh and Colbert, Ian and Cano, Jos\'{e}},
title = {Exploiting Unstructured Sparsity in Fully Homomorphic Encrypted {DNNs}},
year = {2025},
isbn = {9798400715389},
publisher = {ACM},
address = {New York, NY, USA},
doi = {10.1145/3721146.3721948},
booktitle = {Proceedings of the 5th Workshop on Machine Learning and Systems},
pages = {31–38},
numpages = {8},
location = {World Trade Center, Rotterdam, Netherlands},
series = {EuroMLSys '25}
}

@InProceedings{AntonyNaduvath2022CirculantCompletion,
author="Antony, Toby B.
and Naduvath, Sudev",
title="On Circulant Completion of Graphs",
booktitle="Data Science and Security",
year="2022",
publisher="Springer Nature Singapore",
address="Singapore",
pages="181--188",
isbn="978-981-19-2211-4",
doi = {10.1007/978-981-19-2211-4_15}
}

@article{RichterRocha2017,
  author       = {Sebastian Richter and Israel Rocha},
  title        = {Layout of random circulant graphs},
  journal      = {Linear Algebra and its Applications},
  volume       = {559},
  pages        = {95--113},
  year         = {2018},
  doi          = {10.1016/j.laa.2018.09.003},
}

@InProceedings{ParkCCMM2025,
author="Park, Jai Hyun",
editor="Fehr, Serge
and Fouque, Pierre-Alain",
title="Ciphertext-Ciphertext Matrix Multiplication: Fast for Large Matrices",
booktitle="Advances in Cryptology -- EUROCRYPT 2025",
year="2025",
publisher="Springer Nature Switzerland",
address="Cham",
pages="153--180",
isbn="978-3-031-91101-9",
doi={10.1007/978-3-031-91101-9_6}
}

@article{Huang2023,
    author = {Huang, H. and Zong, H.},
    title = {Secure matrix multiplication Based on Fully Homomorphic Encryption.},
    journal = {The Journal of Supercomputing},
    volume = {79},
    pages = {5064–5085},
    doi = {10.1007/s11227-022-04850-4},
    year = {2023}
}

@misc{ChenBicyclic2024,
      author = {Jingwei Chen and Linhan Yang and Wenyuan Wu and Yang Liu and Yong Feng},
      title = {Homomorphic Matrix Operations under Bicyclic Encoding},
      howpublished = {Cryptology {ePrint} Archive, Paper 2024/1762},
      year = {2024},
      doi = {10.1109/TIFS.2024.3490862},
}

@misc{Mahon2025,
      author = {Hannah Mahon and Shane Kosieradzki},
      title = {Encrypted Matrix Multiplication Using 3-Dimensional Rotations},
      howpublished = {Cryptology {ePrint} Archive, Paper 2025/1367},
      year = {2025},
      url = {https://eprint.iacr.org/2025/1367}
}

@inproceedings{Jiang2018,
    author ={Xiaoqian Jiang, Miran Kim and Kristin Lauter and Yongsoo Song},
    title = {Secure Outsourced Matrix Computation and Application to Neural Networks},
    booktitle = {CCS '18: Proceedings of the 2018 ACM SIGSAC Conference on Computer and Communications Security},
    year = {2018},
    doi = {10.1145/3243734.3243837}
}

@inproceedings{Rizomiliotis2022,
author = {Rizomiliotis, Panagiotis and Triakosia, Aikaterini},
title = {On Matrix Multiplication with Homomorphic Encryption},
year = {2022},
isbn = {9781450398756},
publisher = {ACM},
address = {New York, NY, USA},
doi = {10.1145/3560810.3564267},
booktitle = {Proceedings of the 2022 on Cloud Computing Security Workshop},
pages = {53–61},
numpages = {9},
location = {Los Angeles, CA, USA},
series = {CCSW'22}
}

@article{suitesparse,
  author    = {Timothy A. Davis and Yifan Hu},
  title     = {{The University of Florida Sparse Matrix Collection}},
  journal   = {ACM Transactions on Mathematical Software},
  volume    = {38},
  number    = {1},
  pages     = {1:1--1:25},
  year      = {2011},
  month     = {dec},
  doi       = {10.1145/2049662.2049663},
}

@ARTICLE{JoonWoo2022,
  author={Lee, Joon-Woo and Kang, Hyungchul and Lee, Yongwoo and Choi, Woosuk and Eom, Jieun and Deryabin, Maxim and Lee, Eunsang and Lee, Junghyun and Yoo, Donghoon and Kim, Young-Sik and No, Jong-Seon},
  journal={IEEE Access}, 
  title={Privacy-Preserving Machine Learning With Fully Homomorphic Encryption for Deep Neural Network}, 
  year={2022},
  volume={10},
  number={},
  pages={30039-30054},
  doi={10.1109/ACCESS.2022.3159694}}

@ARTICLE{Meftah2021,
  author={Meftah, Souhail and Tan, Benjamin Hong Meng and Mun, Chan Fook and Aung, Khin Mi Mi and Veeravalli, Bharadwaj and Chandrasekhar, Vijay},
  journal={IEEE Transactions on Information Forensics and Security}, 
  title={DOReN: Toward Efficient Deep Convolutional Neural Networks with Fully Homomorphic Encryption}, 
  year={2021},
  volume={16},
  number={},
  pages={3740-3752},
  doi={10.1109/TIFS.2021.3090959}}

@INPROCEEDINGS{williams2007optimization,
  author={Williams, Samuel and Oliker, Leonid and Vuduc, Richard and Shalf, John and Yelick, Katherine and Demmel, James},
  booktitle={SC '07: Proceedings of the 2007 ACM/IEEE Conference on Supercomputing}, 
  title={Optimization of sparse matrix-vector multiplication on emerging multicore platforms}, 
  year={2007},
  volume={},
  number={},
  pages={1-12},
  doi={10.1145/1362622.1362674}}

@INPROCEEDINGS{xu2010sparse,
  author={Xu, Shiming and Lin, Hai Xiang and Xue, Wei},
  booktitle={2010 Ninth International Symposium on Distributed Computing and Applications to Business, Engineering and Science}, 
  title={Sparse Matrix-Vector Multiplication Optimizations based on Matrix Bandwidth Reduction using NVIDIA CUDA}, 
  year={2010},
  volume={},
  number={},
  pages={609-614},
  doi={10.1109/DCABES.2010.162}}

@inproceedings{cryptogcn2022,
author = {Ran, Ran and Xu, Nuo and Wang, Wei and Quan, Gang and Yin, Jieming and Wen, Wujie},
title = {{CryptoGCN}: fast and scalable homomorphically encrypted graph convolutional network inference},
year = {2022},
isbn = {9781713871088},
publisher = {Curran Associates Inc.},
address = {Red Hook, NY, USA},
articleno = {2731},
numpages = {14},
location = {New Orleans, LA, USA},
series = {NIPS '22},
doi = {10.5555/3600270.3603001}
}
\end{document}